\documentclass[aps,prl,twocolumn,amssymb,amsfonts,floatfix,superscriptaddress,longbibliography]{revtex4-2}
\usepackage{epsfig}
\usepackage[section]{placeins}
\usepackage{amsmath}
\usepackage{bm}
\usepackage{longtable}
\usepackage{times}
\usepackage{amssymb}

\newcommand{\beq}{\begin{equation}}
\newcommand{\eeq}{\end{equation}}
\newcommand{\bea}{\begin{eqnarray}}
\newcommand{\eea}{\end{eqnarray}}

\usepackage{color}

\usepackage{ulem} 

\begin{document}
\title{Qubit decoherence under two-axis coupling to low-frequency noises}
\author{Guy Ramon}
\email{gramon@scu.edu}
\affiliation{Department of Physics, Santa Clara University, Santa Clara, CA 95053}
\author{{\L}ukasz Cywi{\'n}ski}
\email{lcyw@ifpan.edu.pl}
\affiliation{Institute of Physics, Polish Academy of Sciences, Aleja Lotnikow 32/46, PL-02668 Warsaw, Poland}

\begin{abstract}

Many solid-state qubit systems are afflicted by low frequency noise mechanisms that operate along two perpendicular axes of the Bloch sphere. Depending on the qubit's control fields, either noise can be longitudinal or transverse to the qubit's quantization axis, thus affecting its dynamics in distinct ways, generally contributing to decoherence that goes beyond pure dephasing. Here we present a theory that provides a unified platform to study dynamics of a qubit subjected to two perpendicular low-frequency noises (assumed to be Gaussian and uncorrelated) under dynamical decoupling pulse sequences. The theory is demonstrated by the commonly encountered case of power-law noise spectra, where approximate analytical results can be obtained.
\end{abstract}

\maketitle

\textit{Introduction.}
Decoherence of qubits can be calculated relatively easily in two cases: that of pure dephasing due to Gaussian longitudinal noise acting along the qubit's energy quantization axis \cite{Paladino_RMP14,Szankowski_JPCM17}, and that of Markovian evolution of open systems that applies when the relevant environmental fluctuations (coupled along any axis) act on timescales shorter than that of the open system dynamics of the qubit \cite{Breuer,Rivas_book}. However, many solid-state qubits decohere due to environmental fluctuations with non-negligible correlation times that couple along at least two perpendicular axes. Born-Markov treatment of both relaxation and dephasing is then inapplicable, and a general solution beyond the pure dephasing case is out of reach \cite{Aihara_PRA90,Risken_PRA90}. In the often-encountered case of noises with spectra concentrated at low frequencies - quasistatic or $1/f$ type \cite{Paladino_RMP14} - an adiabatic treatment of qubit dynamics caused by multi-axis noise is possible \cite{Falci_PRL05,Barnes_PRB16}.
Our focus here is on two-axis coupling of a qubit to such low-frequency noises, and we develop an approximate analytical solution to decoherence for qubit that is freely evolving or subjected to dynamical decoupling (DD) sequences \cite{Viola_PRL99,Kofman_PRL04,Biercuk_JPB11,Lidar_ACP14,Suter_RMP16,Szankowski_JPCM17,Degen_RMP17}.

The Hamiltonian of the qubit-environment system can be written quite generally as:
\begin{equation}
{\cal H}(t)=\frac{1}{2}\left[ \mathbf{B} +{\bm \xi}(t) \right] \cdot {\bm \sigma},
\label{H}
\end{equation}
where $\mathbf{B}$ is a vector of the qubit control fields, ${\bm \xi}(t)$ is a vector of environmental quantum operators or classical stochastic functions representing noise, and ${\bm \sigma}$ is the vector of Pauli matrices. Strictly speaking, the qubit control fields are not static, as they typically include dynamical decoupling (DD) pulse sequences, but
here we assume instantaneous pulses that result in perfect $\pi$ rotations of the qubit state around the $y$ axis, perpendicular to both control and noise directions.

Solid state devices are abundant with sources of low-frequency excitations such as slowly switching two-level fluctuators responsible for $1/f$ noise \cite{Paladino_RMP14}.
Prominent examples of solid-state based qubits affected by two-axis low-frequency noise include those based on two \cite{Petta_Science05,Taylor_NatPhys05,Taylor_PRB07} or three \cite{Medford_PRL13,Medford_NN13,Russ_PRB16,Sala_PRB17,Russ_PRL18}
exchange-coupled semiconductor quantum dots (QDs) containing at least two electrons, and both charge and flux superconducting (SC) qubits \cite{Bylander_NatPhys11,Yan_NatComm16,Krantz_APR19}. In all these devices, electronic charge noise and flux noise spectra follow power law, $1/f^{\alpha}$, over a wide range of frequencies, with $\alpha$ generally falling in the range of $\alpha =1 - 1.25$ \cite{Nakamura_PRL02,Astafiev_PRL04,Jung_APL04,Jung_APL04,Gustafsson_PRB13,Freeman_APL16}.
Several experiments reported other power laws, including $\alpha\! = \! 0.9$ for flux noise in a SC qubit \cite{Bylander_NatPhys11}, $\alpha=0.7$ for charge noise in GaAs QDs \cite{Dial_PRL13}, $\alpha=1.93$ in a charge-tunable SC device inflicted with anomalous large-amplitude charge noise \cite{Christensen_PRB19}, and a dual power law of $\alpha=1.48/1.97$ of charge noise in Si QD, where the higher power law was measured at extremely low frequencies, below $10^{-4}$ Hz \cite{Struck_npjqi20}.


We focus here on the case of two-electron singlet-triplet ($S-T_0$) qubit in a double quantum dot (DQD),
for which $\mathbf{B} =(\delta h,0,J)$, where $\delta h$ is the interdot magnetic field gradient across the QDs and $J$ is the exchange coupling \cite{Coish_PRB05,Taylor_PRB07}. The latter originates from Coulomb interaction, and as such exhibits slow fluctuations \cite{Hu_PRL06,Culcer_APL09,Ramon_PRB10,Ramon_PRB12,Dial_PRL13} caused by $1/f^{\alpha}$ charge noise.
Finite $\delta h$ arises due to a spatially dependent field from a nanomagnet \cite{Brunner_PRL11,Yoneda_NN18,Zajac_Science19,Struck_npjqi20}
or inhomogeneous nuclear spin polarization resulting in Overhauser field gradient \cite{Foletti_NatPhys09,Bluhm_PRL10}.
In the latter case, nuclear noise is concentrated at very low frequencies \cite{Reilly_PRL08,Malinowski_PRL17}, and the quasi-static approximation breaks down only at timescales longer than $10\mu\mathrm{s}$ \cite{Witzel_PRB06,Witzel_PRL07,Bluhm_NatPhys11}. However, charge noise leads to stochastic shifts of the electron wavefunctions with respect to the frozen nuclei, thus making the Overhauser fields experienced by the electrons inherit the characteristics of charge noise \cite{Bluhm_NatPhys11,Malinowski_PRL17}. The same happens when $\delta h$ results from an external magnetic field gradient: charge noise induces variations in electron positions that translate into fluctuations of their spin splitting, and thus $\delta h$. Consequently, noise in both $\delta h$ and $J$ is of $1/f^{\alpha}$ type at high frequencies, with an additional zero-frequency component for $\delta h$ accounting for nuclear spin diffusion. In GaAs QDs, $\delta h$  noise power spectra characterized by $\alpha=1-2.6$ were measured at frequencies between $\sim \! 1\,\mathrm{kHz}$, below which classical nuclear spin diffusion results in a Lorentzian, quasi-static noise, and $\sim \! 100\, \mathrm{kHz}$ \cite{Reilly_PRL08,Medford_PRL12,Malinowski_PRL17,Malinowski_NatNano17}.
It should be stressed that our model for two low-frequency noises applies to all the above-mentioned qubits, so while we present below results for the $S$-$T_0$ qubit, our theory  applies to a wide class of systems.

Energy relaxation of the qubit depends on the availability of environmental excitations with appreciable transverse (with respect to the quantization axis set by $\mathbf{B}$) coupling to the qubit and energy that is comparable to its level splitting. In contrast, environmental degrees of freedom with any energy contribute to pure dephasing of superpositions of qubit's eigenstates. In devices with strong low-frequency noises, the timescales of dephasing and relaxation are thus often well-separated, with coherence becoming limited by relaxation only after application of a very large number of DD pulses \cite{Bylander_NatPhys11}. This justifies our neglect of relaxation, and focus on effects of dephasing and tilting of the qubit's quantization axis. A crucial element of our theory follows from the fact that transverse noise
couples to the qubit's phase nonlinearly (quadratically in the lowest order). As a result, even a noise with Gaussian statistics becomes effectively non-Gaussian and calculation of its higher order cumulants, beyond the second one, is necessary to correctly evaluate the qubit's dephasing \cite{Makhlin_PRL04,Falci_PRL05,Bergli_PRB06,Cywinski_PRA14}.

In this work we develop a unified theory for the time evolution of a qubit state under two uncorrelated, zero-mean, low-frequency Gaussian noises that operate on perpendicular axes \cite{correlations}. Our theory extends a previous analysis made by Barnes {\it et al.} \cite{Barnes_PRB16} for $S$-$T_0$ qubits in two respects: (i) we perform the calculation to second order in ${\bm \xi}(t)$, such that (quadratically coupled) transverse noise is considered, and (ii) we include DD control pulse sequences, accounting for qubit evolution outside the free induction decay (FID) case. The latter is made possible by the former, as effects of transverse noise are typically overshadowed by the longitudinal one when no DD filtering of the lowest-frequency longitudinal noise is done. We also include contributions resulting from the nontrivial interplay of longitudinal and transverse noises, as well as axis-tilting effects, thus providing a complete analytical treatment of the problem of decohrrence due to two-axis slow noises.


\textit{Formalism.} We specify the qubit working position for a two-axis control field, ${\mathbf B}=(B_x, 0, B_z)$, using the angle $\overline{\chi}=\arctan (B_x/B_z)$.
$\xi_z$, $\xi_x$ in the Hamiltonian, Eq.~(\ref{H}), represent fluctuations of the respective control fields.
We assume that these act on a much slower timescale, as compared with the qubit dynamics, allowing us to take the adiabatic limit, where the qubit evolution operator is approximated by applying instantaneous eigenstates of ${\cal H}(t)$ \cite{Barnes_PRB16}. The resulting instantaneous unitary evolution reads
\begin{equation}
U(t) \! \approx \! \left( \!\!
\begin{array}{cc} \cos \phi-i\sin \phi \cos \chi & -i \sin \phi \sin \chi \\ -i \sin \phi \sin \chi & \cos \phi+i \sin \phi \cos \chi
\end{array} \!\! \right),
\end{equation}
where the noises impact the evolution by modifying the rotation axis, $\chi(t)$, and the accumulated rotation angle $\phi(t)$:
\begin{eqnarray}
\chi (t)&=& \arctan \left( \frac{B_x+\xi_x}{B_z+\xi_z} \right) \equiv \overline{\chi} + \delta \chi (t) \\
\phi(t) &=& \overline{\phi}(t) +\delta \phi (t).
\end{eqnarray}
Without noise we have $\overline{\phi}(t)=\frac{1}{2} \int_0^t dt' f_t (t') B_z \sec \overline{\chi}$, where $f_t(t')$ is the switching function corresponding to the employed pulse protocol, whose Fourier transform, $\tilde{f}_t(\omega)$, is known as the filter function \cite{Cywinski_PRB08}. For FID, $\overline{\phi}(t)=\sqrt{B_z^2+B_x^2} t/2$, whereas any balanced pulse protocol yields $\overline{\phi}(t)=0$, since $\int f_t(t')dt' =0$. One can split the qubit-environment term in the Hamiltonian, Eq.~(\ref{H}), into parts that are parallel and perpendicular to the quit control axis, using $\xi_{\parallel}=\xi_x \sin \overline{\chi}+\xi_z \cos \overline{\chi}$ and $\xi_{\perp}=\xi_x \cos \overline{\chi}-\xi_z \sin \overline{\chi}$. To second order in $\xi_l$, $l \in \{\parallel,\perp\}$ we have:
\begin{eqnarray}
\delta \chi(t)& \approx & \frac{\cos \overline{\chi}}{B_z} \xi_\perp (t) \left[1 -\frac{\cos \overline{\chi}}{B_z} \xi_{\parallel}(t)\right] \label{delchi} \\
\delta \phi (t) & \approx & \frac{1}{2} \int_0^t dt' f_t(t') \left[\xi_{\parallel} (t') + \frac{\cos \overline{\chi}}{2B_z} \xi_{\perp}^2(t') \right].
\label{delphi}
\end{eqnarray}

Utilizing the qubit Hamiltonian eigenstates in the tilted rotation axis $z'=(\sin \overline{\chi},0,\cos \overline{\chi})$:
\begin{equation}
| + \rangle= \left( \begin{array}{c} \cos \frac{\overline{\chi}}{2} \\ \sin \frac{\overline{\chi}}{2} \end{array} \right), \hspace{0.5 cm}
| - \rangle= \left( \begin{array}{c} -\sin \frac{ \overline{\chi}}{2} \\ \cos \frac{\overline{\chi}}{2} \end{array} \right),
\end{equation}
and the perpendicular state: $|x' \rangle = \frac{1}{2}\left(|+\rangle +| - \rangle \right)$, the effects of the two noises can be quantified by the coherence function:
\begin{eqnarray}
W(t) &=& \frac{\left| \langle \rho_{+-} (t) \rangle \right|}{\left| \langle \rho_{+-}(0) \rangle \right|}= \left| \! \left \langle \frac{\langle + | U(t)| x'\rangle \langle x'| U^\dagger (t)| - \rangle}{\langle + | x'\rangle \langle x'| +\rangle} \right\rangle \! \right| = \notag \\
&& \left| \langle 1-2\sin\phi \cos \delta \chi \left( \cos \delta \chi \sin \phi +i\cos \phi \right) \rangle \right| \approx \notag \\
&& \left|e^{-2i \overline{\phi}} \left\langle e^{-2i\delta \phi} \right\rangle -\frac{1}{2} \left\langle \delta \chi^2 \left[ \cos 2\phi + e^{-2i \phi} \right] \right\rangle \right|,
\label{W}
\end{eqnarray}
where $\langle \cdot \rangle$ denotes Gaussian averaging over both $\xi_z$ and $\xi_x$, and the last row is correct to second order in these noises, with the first (second) term corresponding to rotation angle (axis tilting) error. 

The presence of quadratic noise terms in $\delta \phi$ requires a full cumulant expansion in the averaging since $\xi_l^2(t)$ are no longer Gaussian distributed \cite{Cywinski_PRA14}. For zero-mean Gaussian noises, $\langle \xi^k (t) \rangle =0$ for odd $k$, and even-power terms factorize to two-point correlators, $\langle \xi(t_1) \xi(t_2) \rangle \equiv S(t_{12})$, where $t_{12} \equiv t_1-t_2$. Addressing first the dominant contribution due to rotation angle errors we have
\begin{equation}
\langle e^{\pm 2i \delta \phi} \rangle = \exp \left\{{\sum_{k=1} (\pm i)^k \frac{C_k}{k!}}\right\},
\label{dominant}
\end{equation}
where $C_k$ generalize the standard noise cumulants \cite{Kubo} for two uncorrelated noises, and are given explicitly in terms of their noise power spectra in section I of the supplemental material \cite{SM}. 

The structure of the $k$th cumulant reveals two types of contributions that we coin {\it linked}, $R_k(t)$, (with $k$ correlators) and {\it semi-linked}, $\tilde{R}_k(t)$, (with $k-1$ correlators):
\begin{eqnarray}
R_{k}(t) &\!=\!& -\frac{1}{2k}\! \left( \frac{i}{B} \right)^{k} \!\!\! \int_0^{\infty} \!\! \frac{d\omega_1 \cdots d \omega_{k}}{\pi^{k}} \tilde{f}_t(\omega_{12}) \cdots  \tilde{f}_t(\omega_{k1}) \notag \\
&& \prod_{i=1}^{k} \left[ \sin^2 \overline{\chi} \tilde{S}_z(\omega_i)\!+\cos^2\overline{\chi} \tilde{S}_x(\omega_i) \right],
\label{R}
\end{eqnarray}
\begin{eqnarray}
\tilde{R}_{k}(t) &\!=\!& -\frac{1}{2}\! \left( \frac{i}{B} \right)^{k}\! (B_z\sin \overline{\chi})^2 \! \int_0^{\infty} \! \frac{d\omega_1 \cdots d \omega_{k-1}}{\pi^{k-1}} \times \notag \\
&& \tilde{f}_t(-\omega_1) \left[ \tilde{S}_z(\omega_1)- \tilde{S}_x(\omega_1)\right] \tilde{f}_t(\omega_{12})\times  \notag \\
&& \prod_{i=2}^{k-2} \tilde{f}_t(\omega_{i,i+1}) \left[ \sin^2 \overline{\chi} \tilde{S}_z(\omega_i)\!+\cos^2\overline{\chi} \tilde{S}_x(\omega_i) \right] \times \notag \\
&&\tilde{f}_t(\omega_{k-1}) \left[\tilde{S}_z(\omega_{k-1})\!- \tilde{S}_x(\omega_{k-1}) \right]\!.
\label{Rt}
\end{eqnarray}
In Eqs.~(\ref{R})-(\ref{Rt}), $\tilde{S}_z(\omega)$ and $\tilde{S}_x(\omega)$ are the power spectra of the two noises. 
The linked diagrams involve only $\xi_{\perp}^2$ contributions, whereas the semi-linked diagrams include a mixing of $\xi_{\perp}^2$ and $\xi_{\parallel}$ terms. Eq.~(\ref{dominant}) then reads
\begin{equation}
\langle e^{\pm 2i \delta \phi} \rangle = c_z(t) c_x(t) e^{-\left[\Sigma_{2k}+\tilde{\Sigma}_{2k}\right]} e^{\pm i\left[\Sigma_{2k+1}+\tilde{\Sigma}_{2k+1}\right]},
\label{I}
\end{equation}
where $\Sigma_{2k} \! \equiv \! \sum_{k=1} R_{2k} (t)$ $\left[\tilde{\Sigma}_{2k} \! \equiv \! \sum_{k=2} \tilde{R}_{2k}(t)\right]$ and $\Sigma_{2k+1} \! \equiv \! i\sum_{k=0} R_{2k+1} (t)$ $\left[\tilde{\Sigma}_{2k+1} \! \equiv \! i\sum_{k=1} \tilde{R}_{2k+1}(t)\right]$ are the summations over linked [semi-linked] even and odd diagrams, respectively, and we singled out the semi-linked contributions in the second cumulant that are accounted for in \cite{Barnes_PRB16}:
\begin{eqnarray}
c_z(t) \!&\!=\!&\! \exp{\left\{-\cos^2 \overline{\chi} \int_0^{\infty} \frac{d\omega}{2\pi} |\tilde{f}_t(\omega) |^2 \tilde{S}_z(\omega)\right\}} \notag \\
c_x(t) \!&\!=\!&\! \exp{\left\{-\sin^2 \overline{\chi} \int_0^{\infty} \frac{d\omega}{2\pi} |\tilde{f}_t(\omega) |^2 \tilde{S}_x(\omega)\right\}},
\label{CJCh}
\end{eqnarray}
For odd number of DD pulses, $\tilde{f}_t (\omega)$ is an odd function and only even cumulants survive. In this case the phase terms in Eq.~(\ref{I}) vanish and only signal decay remains.

The evaluation of the axis-error, transient, contribution in Eq.~(\ref{W}) is more involved and the calculational details can be found in section II of the supplemental material \cite{SM}. We note here that the leading terms in this contribution vanish for any balanced DD pulse sequence. Finally, we provide for completeness, formulas for singlet and $| \uparrow \downarrow \rangle$ return probabilities, correct to second order in noise amplitudes, under any DD pulse sequence \cite{SM}, corresponding to experiments reported in Refs.~\cite{Malinowski_PRL17} and \cite{Martins_PRL16}, respectively.

\textit{Cumulant resummation for low-frequency noises.}
We use the resummation technique of ref.~\cite{Cywinski_PRA14} to derive analytical results for the cumulant sums found in the rotation angle error contribution, Eq.~(\ref{I}), and for the various time derivatives of these sums in the axis-error contribution \cite{SM}.
We split the noise into a dominant low-frequency, static component and a high-frequency, time-dependent component:
$ \xi_{l}=\xi_{l}^{\rm lf}+\xi_{l}^{\rm hf}(t)$, and denote the standard deviations of the low- and high-frequency noise components as:
\begin{eqnarray}
\sigma_{0l}^2 = \int_{\omega_0}^{\omega_1} \frac{d\omega}{\pi} \tilde{S}_{l} (\omega)  \,\, , \,\, 
\sigma_{tl}^2 = \int^{\infty}_{\omega_1} \frac{d\omega}{\pi} \tilde{S}_{l} (\omega),
\end{eqnarray}
where $\omega_0$ is a low-frequency cutoff, determined by the shorter of the noise correlation time and the total acquisition time, both of which are typically much longer than $t$, and $\omega_1$ can be taken as $1/t$ or otherwise as a fixed ultraviolet cutoff. Our approximate calculation of the cumulant sums rests on the assumption that $\sigma_{tl}^2 \ll \sigma_{0l}^2$, for both noises at timescales relevant for the qubit operation. This assumption holds for any power-law noise with $\alpha\geq 1$.

Replacing each noise correlator with $S_{l}(t_{ij})=\langle \xi^{\rm hf}_{l} (t_i) \xi^{\rm hf}_{l}(t_j)\rangle_{\rm hf}+\sigma_{0l}^2$, and keeping only terms with maximal power of $\sigma_{0l}^2$, we derive explicit expressions for the cumulant terms and their time derivatives \cite{SM}, and perform the summations in Eq.~(\ref{I}). Whereas for any balanced DD sequence this procedure amounts to replacing every second correlator with $\sigma_{0l}^2$, in the FID case \textit{\emph{}all} correlators are replaced with $\sigma_{0l}^2$. The linked and semi-linked even sums are found respectively as
\begin{equation}
e^{-\Sigma_{2k}} = \left\{ \begin{array}{ll} \sqrt{\eta (t)}, & \mbox{DD} \\ \eta_{\rm FID}^{1/4} (t), & \mbox{FID} \end{array} \right.
\label{R2klf}
\end{equation}
and
\begin{equation}
\tilde{\Sigma}_{2k} \!=\! \left\{ \!\! \begin{array}{ll}\! \frac{\eta(t)}{2}\!\! \left(\!\!\frac{B_x B_z}{B^3}\!\right)^2\!\!  \left[ \tilde{f}^2_t (0) \overline{S}_+^{\rm hf}(t) \sigma_{0-}^4\!+\!\left[S_-^{\rm hf} (t) \right]^2 \overline{\sigma}_{0+}^2\right]\!, & \!\! \rm{ DD} \\ \frac{\eta_{\rm FID}(t)}{2} \left(\!\!\frac{B_x B_z}{B^3}\!\right)^2 \sigma_{0-}^4 \overline{\sigma}_{0+}^2 t^4, & \!\!{\rm FID} \end{array} \right.
\label{R2ktlf}
\end{equation}
where we have defined
\begin{equation}
\eta (t) \!\equiv \!\left(\! 1\!+\! \frac{\overline{\sigma}_{0+}^2 \overline{S}_+^{\rm hf} (t)}{B^2} \!\right)^{-1}\!\!\!; \hspace{0.05 cm} \eta_{\rm FID} (t) \!\equiv \!\left( \! 1 \!+\!\frac{\overline{\sigma}_{0+}^4 t^2}{B^2}\right)^{-1}\!\! .
\label{eta}
\end{equation}
In Eqs.~(\ref{R2ktlf})-(\ref{eta}), $\sigma^2_{0\pm} \equiv \sigma^2_{0z} \pm \sigma^2_{0x}$, $\overline{\sigma}^2_{0\pm} \equiv\sin^2 \overline{\chi} \sigma^2_{0z} \pm \cos^2 \overline{\chi} \sigma^2_{0x}$, and similarly the high-frequency combined noise correlators are given by $S_\pm^{\rm hf} (t) = S_{z}^{\rm hf} (t) \pm S_{x}^{\rm hf} (t)$, $\overline{S}_\pm^{\rm hf} (t) = \sin^2 \overline{\chi} S_{z}^{\rm hf} (t) \pm \cos^2 \overline{\chi} S_{x}^{\rm hf} (t)$, where $S_{l}^{\rm hf}(t)=\int_{\omega_1}^\infty \frac{d\omega}{\pi} \left|\tilde{f}_t(\omega)\right|^2 \tilde{S}_{l}(\omega)$. The sums over odd diagrams in Eq.~(\ref{I}) are nonzero only for FID, giving a nontrivial phase shift in $W(t)$ that is characteristic for free evolution dephasing due to low-frequency transverse noise \cite{Falci_PRL05,Koppens_PRL07,Cywinski_PRB09}:
\begin{eqnarray}
\Sigma_{2k+1}^{\rm FID} (t) &\!=\!& \frac{1}{2} \arctan \left( \frac{\overline{\sigma}_{0+}^2 t}{B} \right)\notag \\
\tilde{\Sigma}_{2k+1}^{\rm FID}(t) &\!=\!& -\frac{\sin^2 2\overline{\chi}}{8B} \eta_{\rm FID}(t) \sigma_{0-}^4 t^3.
\label{R2kp1lf}
\end{eqnarray}
We calculated the highest subleading contribution due to linked odd diagrams, with one less $\sigma_{0l}^2$ factor, showing it to be negligible for experimentally relevant noise parameters \cite{SM}.

\textit{Results.}
We now demonstrate the versatility of our two-axis noise theory in predicting decoherence at arbitrary working positions, by considering real-life noise parameters, pertaining to the charge ($J$) and magnetic ($H$) control fields in singlet-triplet spin qubits. We consider $\alpha_J=\alpha_H=1$, such that $\tilde{S}_{J/H} =A_{J/H}^2/\omega$ for both noise spectra with a low-frequency cutoff of $\omega_0=1$ Hz, and also include for the nuclear noise a quasi-static contribution $S^{qs}_{H} \! =\! \sigma^2_{0H}\delta(\omega)$. Unless otherwise noted, we take the high-frequency nuclear noise amplitude as $A_H=66$ peV \cite{Malinowski_NatNano17}, attributed to shaking of the electronic wavefunction by charge noise. For $\alpha_J\!=\! 1$ we have $A_J \! \approx \sigma_{0J}/5$ at typical $\omega_0$ values, and $A_{J}\! \approx 10^{-3}J$, as was measured in \cite{Dial_PRL13}.

\begin{figure}[tb]
\epsfxsize=1\columnwidth
\vspace*{-0.9 cm}
\centerline{\epsffile{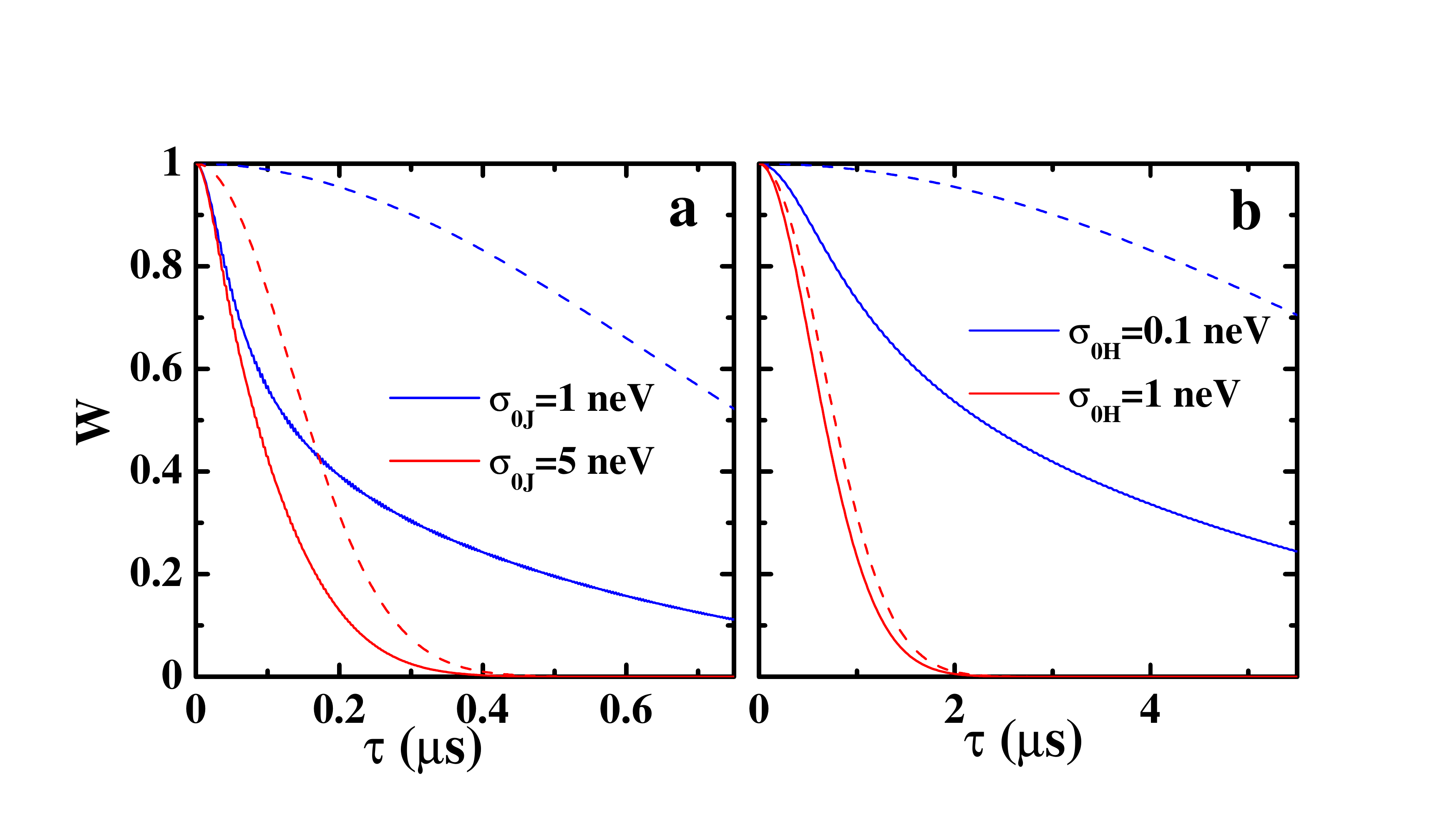}}
\vspace*{-0.6 cm}
\caption{FID decoherence function, Eq.~(\ref{W}), calculated to first order (dashed lines) and with a full cumulant summation (solid lines). (a) $J=0.5 \mu$eV, $\delta h=0$ ($\overline{\chi}=0$), $\sigma_{0H}=0.1 \mu $eV relevant for GaAs QDs, $\sigma_{0J}= 1$ neV (Blue lines) and 5 neV (Red Lines), and $A_J=\sigma_{0J}/5$. The perpendicular noise contribution becomes dominant with smaller charge noise amplitudes. (b) $J=0$, $\delta h= 0.1 \mu$eV ($\overline{\chi}=\pi/2$), $\sigma_{0J}=10\mathrm{neV}$, and $A_J=0.2 \sigma_{0J}$. We consider low static magnetic noise values $\sigma_{0H}=1$ neV, $0.1$ neV with the latter relevant for isotopically purified Si QDs with micromagnets \cite{Yoneda_NN18,Struck_npjqi20}.
}
\label{Fig1}
\end{figure}

In Figs.~\ref{Fig1}a and b we consider FID at $\overline{\chi}=0$, and $\pi/2$, respectively, focusing on scenarios where the transverse noise contribution is comparable or greater than the longitudinal one. In either case the main contribution to the longitudinal noise comes from $c_{x/z}$, Eq.~(\ref{CJCh}), resulting in dephasing time of $T_{2 \|}=\sqrt{2}/\sigma_{0 \|}$, whereas the dominant contribution to the transverse noise comes from the linked terms, Eqs.~(\ref{R2klf}), (\ref{eta}), resulting in dephasing time of $T_{2\perp} \approx 7.3B_{\|}/\sigma_{0 \perp}^2$ \cite{semilinked}. For $\overline{\chi}=0$ the transverse (nuclear) noise contribution can easily dominate dephasing due to the relatively large Overhauser field gradient static noise of $\sigma_{0H}=0.1 \mu $eV, measured for GaAs QDs \cite{Malinowski_NatNano17} (see Fig.~\ref{Fig1}a), but at $\overline{\chi}=\pi/2$, the transverse (charge) noise contribution becomes important only for quiet magnetic environment, e.g., by implementing a field gradient with local micromagnets ($\sigma_{0H} \lesssim 0.1\,\mathrm{neV}$ was measured in isotopically purified Si QDs with nanomagnets and charge noise dominating spin dephasing \cite{Yoneda_NN18,Struck_npjqi20}.)
%
%
As the quantization axis tilts, $\overline{\chi}\gtrsim 0$, the longitudinal noise contribution includes nuclear noise component, thus becoming dominant with a resulting Gaussian decay. This is demonstrated in Fig.~2a, where we provide $T_2$ FID times vs.~$\delta h$ for $J=0.5 \mu$eV. With increasing $\delta h$, decay is dominated by longitudinal contribution, adequately described by the first order calculation.

In order to provide an intuitive explanation for decoherence at DD setting, we consider quasi-static nuclear noise ($A_H=0$) and again limit our discussion to the linked terms, Eqs.~(\ref{R2klf}), (\ref{eta}) (this picture is largely unchanged if small dynamic nuclear noise is added). Starting at $\overline{\chi}=0$ we have
Gaussian decay due to longitudinal noise, $T_{2J} \approx 3/A_J$, while the quasi-static transverse noise is echoed away.
As $\overline{\chi}$ increases the longitudinal dephasing time becomes $T_{2\|} = T_{2J}/\cos\overline{\chi}$, whereas under the reasonable assumptions $A_J \gg A_H$, $\sigma_{0J} \ll \sigma_{0H}$, Eq.~(\ref{eta}) gives $T_{2\perp} \approx 2 B T_{2J}/ (\sigma_{0H} \sin 2\overline{\chi})$, as long as $\tan \overline{\chi} \ll \sigma_{0H}/\sigma_{0J}$ is fulfilled. As $\overline{\chi}$ approaches $\pi/2$, longitudinal noise becomes irrelevant whereas the transverse dephasing time saturates at
$T_{2\perp} \approx B T_{2J}/\sigma_{0J}$. Fig.~2b illustrates this nontrivial two-axis behavior, showing a crossover from power-law to Gaussian decay for SE.
\begin{figure}[tb]
\epsfxsize=1\columnwidth
\vspace*{-0.9 cm}
\centerline{\epsffile{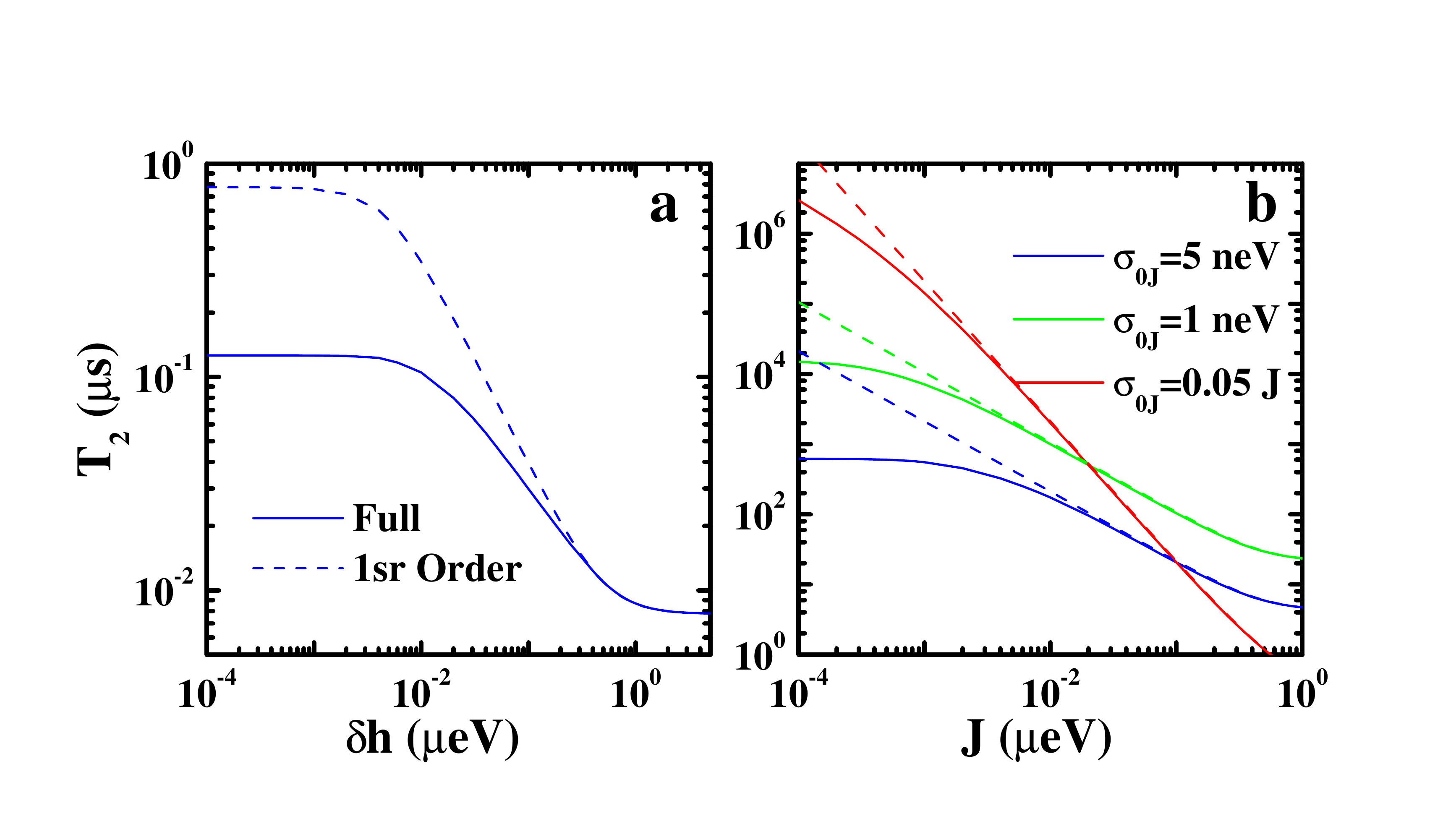}}
\vspace*{-0.7 cm}
\caption{Dephasing times for quasi-static nuclear noise ($\sigma_{0H}=0.1\mu$eV, $A_H=0$) as a function of (a) $\delta h$ for FID at $J=0.5\mu$eV, $\sigma_{0J}= 1$ neV and (b) $J$ for SE at $\delta h=0.5 \mu$eV. Several charge noise models are shown in (b), including $\sigma_{0J}=5$ neV (red), 1 neV (green) and $0.05J$ (blue). Solid lines depict full cumulant summation and dashed lines show first order calculation. In all cases, $A_J=\sigma_{0J}/5$.}
\label{Fig2}
\end{figure}

Finally, Eq.~(\ref{eta}) suggests that the transverse dynamic noise contribution is renormalized by $\overline{\sigma}_{0+}^2$, resulting in an unexpected effect whereby longitudinal quasi-static noise can impact decoherence under DD. This effect is demonstrated in Fig.~3, where we consider $\delta h \gg J$ and show that increasing the (predominantly) longitudinal nuclear quasi-static noise from $\sigma_{0H}=0.01\mu$eV to $0.1 \mu$eV results in $25\%$ ($37\%$) reduction in dephasing time with (without) additional dynamic noise. We note that additional longitudinal-transverse noise mixing results from the semi-linked contributions.

\begin{figure}[tbh]
\epsfxsize=0.7\columnwidth
\vspace*{-0.5 cm}
\centerline{\epsffile{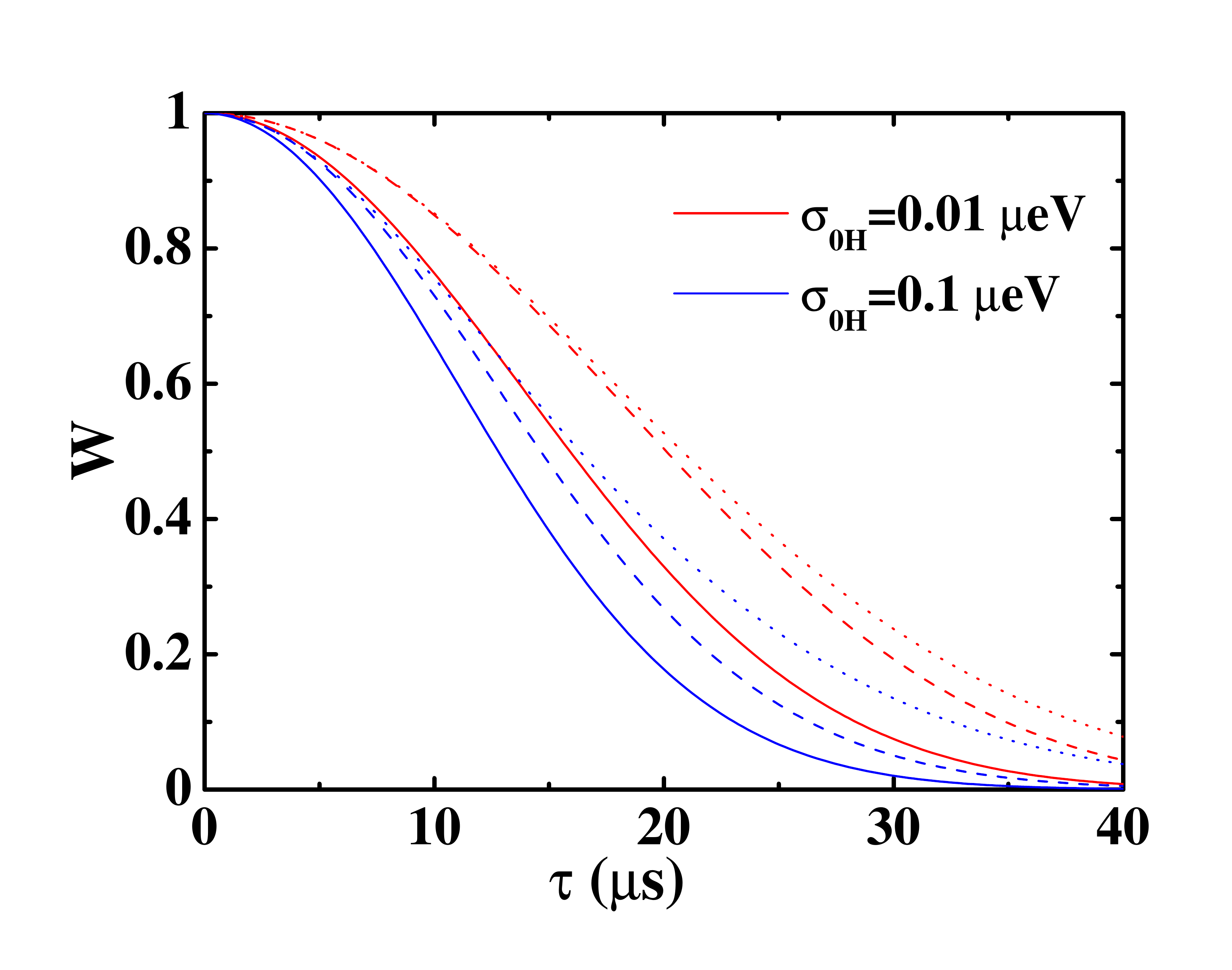}}
\vspace*{-0.5 cm}
\caption{Decoherence function vs.~time for SE at $\delta h=0.1\mu$eV, $J=0.02\mu$eV. The charge noise parameters are $\sigma_{0J}=5$ neV and $A_J=\sigma_{0J}/5$, and we compare nuclear quasi-static noise of $\sigma_{0H}=0.01\mu$eV (dashed red) and $0.1\mu$eV (dashed blue). Solid lines depict decoherence with additional dynamic nuclear noise with amplitude $A_H \!= \! 66$ peV. Dotted lines illustrate the results for $A_{H}\! =\! 0$ excluding the semi-linked contribution from Eq.~(\ref{R2ktlf}).}
\label{Fig3}
\end{figure}

\textit{Conclusions.}
We have developed a theory to evaluate qubit state evolution under two perpendicular low-frequency noises, and obtained closed-form results for the decoherence in both FID and DD settings, by utilizing cumulant summations. Our theory captures the qubit's dynamics at any working point, including the optimal point, where transverse noise (missing in previous first order treatments) dominates and near that point, where the interplay between longitudinal and transverse noises leads to nontrivial dynamics.



\textit{Acknowledgements.} Guy Ramon wishes to thank Yonatan Ramon for working out elements of the calculations detailed in Section IV of the supplemental material \cite{SM}. This work was supported by the National Science Foundation Grant no.~DMR 1829430.

\bibliography{refs_two_axis}

\begin{thebibliography}{61}%
\makeatletter
\providecommand \@ifxundefined [1]{%
 \@ifx{#1\undefined}
}%
\providecommand \@ifnum [1]{%
 \ifnum #1\expandafter \@firstoftwo
 \else \expandafter \@secondoftwo
 \fi
}%
\providecommand \@ifx [1]{%
 \ifx #1\expandafter \@firstoftwo
 \else \expandafter \@secondoftwo
 \fi
}%
\providecommand \natexlab [1]{#1}%
\providecommand \enquote  [1]{``#1''}%
\providecommand \bibnamefont  [1]{#1}%
\providecommand \bibfnamefont [1]{#1}%
\providecommand \citenamefont [1]{#1}%
\providecommand \href@noop [0]{\@secondoftwo}%
\providecommand \href [0]{\begingroup \@sanitize@url \@href}%
\providecommand \@href[1]{\@@startlink{#1}\@@href}%
\providecommand \@@href[1]{\endgroup#1\@@endlink}%
\providecommand \@sanitize@url [0]{\catcode `\\12\catcode `\$12\catcode
  `\&12\catcode `\#12\catcode `\^12\catcode `\_12\catcode `\%12\relax}%
\providecommand \@@startlink[1]{}%
\providecommand \@@endlink[0]{}%
\providecommand \url  [0]{\begingroup\@sanitize@url \@url }%
\providecommand \@url [1]{\endgroup\@href {#1}{\urlprefix }}%
\providecommand \urlprefix  [0]{URL }%
\providecommand \Eprint [0]{\href }%
\providecommand \doibase [0]{https://doi.org/}%
\providecommand \selectlanguage [0]{\@gobble}%
\providecommand \bibinfo  [0]{\@secondoftwo}%
\providecommand \bibfield  [0]{\@secondoftwo}%
\providecommand \translation [1]{[#1]}%
\providecommand \BibitemOpen [0]{}%
\providecommand \bibitemStop [0]{}%
\providecommand \bibitemNoStop [0]{.\EOS\space}%
\providecommand \EOS [0]{\spacefactor3000\relax}%
\providecommand \BibitemShut  [1]{\csname bibitem#1\endcsname}%
\let\auto@bib@innerbib\@empty
\bibitem [{\citenamefont {Paladino}\ \emph {et~al.}(2014)\citenamefont
  {Paladino}, \citenamefont {Galperin}, \citenamefont {Falci},\ and\
  \citenamefont {Altshuler}}]{Paladino_RMP14}%
  \BibitemOpen
  \bibfield  {author} {\bibinfo {author} {\bibfnamefont {E.}~\bibnamefont
  {Paladino}}, \bibinfo {author} {\bibfnamefont {Y.~M.}\ \bibnamefont
  {Galperin}}, \bibinfo {author} {\bibfnamefont {G.}~\bibnamefont {Falci}},\
  and\ \bibinfo {author} {\bibfnamefont {B.~L.}\ \bibnamefont {Altshuler}},\
  }\bibfield  {title} {\bibinfo {title} {$1/f$ noise: Implications for
  solid-state quantum information},\ }\href
  {https://doi.org/10.1103/RevModPhys.86.361} {\bibfield  {journal} {\bibinfo
  {journal} {Rev. Mod. Phys.}\ }\textbf {\bibinfo {volume} {86}},\ \bibinfo
  {pages} {361} (\bibinfo {year} {2014})}\BibitemShut {NoStop}%
\bibitem [{\citenamefont {Sza\'nkowski}\ \emph {et~al.}(2017)\citenamefont
  {Sza\'nkowski}, \citenamefont {Ramon}, \citenamefont {Krzywda}, \citenamefont
  {Kwiatkowski},\ and\ \citenamefont {Cywi\'nski}}]{Szankowski_JPCM17}%
  \BibitemOpen
  \bibfield  {author} {\bibinfo {author} {\bibfnamefont {P.}~\bibnamefont
  {Sza\'nkowski}}, \bibinfo {author} {\bibfnamefont {G.}~\bibnamefont {Ramon}},
  \bibinfo {author} {\bibfnamefont {J.}~\bibnamefont {Krzywda}}, \bibinfo
  {author} {\bibfnamefont {D.}~\bibnamefont {Kwiatkowski}},\ and\ \bibinfo
  {author} {\bibfnamefont {{\L}.}~\bibnamefont {Cywi\'nski}},\ }\bibfield
  {title} {\bibinfo {title} {Environmental noise spectroscopy with qubits
  subjected to dynamical decoupling},\ }\href
  {https://doi.org/10.1088/1361-648X/aa7648} {\bibfield  {journal} {\bibinfo
  {journal} {J. Phys.:Condens. Matter}\ }\textbf {\bibinfo {volume} {29}},\
  \bibinfo {pages} {333001} (\bibinfo {year} {2017})}\BibitemShut {NoStop}%
\bibitem [{\citenamefont {Breuer}\ and\ \citenamefont
  {Petruccione}(2007)}]{Breuer}%
  \BibitemOpen
  \bibfield  {author} {\bibinfo {author} {\bibfnamefont {H.}~\bibnamefont
  {Breuer}}\ and\ \bibinfo {author} {\bibfnamefont {F.}~\bibnamefont
  {Petruccione}},\ }\href@noop {} {\emph {\bibinfo {title} {The Theory of Open
  Quantum Systems}}}\ (\bibinfo  {publisher} {Oxford University Press},\
  \bibinfo {address} {Oxford},\ \bibinfo {year} {2007})\BibitemShut {NoStop}%
\bibitem [{\citenamefont {Rivas}\ and\ \citenamefont
  {Huelga}(2012)}]{Rivas_book}%
  \BibitemOpen
  \bibfield  {author} {\bibinfo {author} {\bibfnamefont {{\'A}.}~\bibnamefont
  {Rivas}}\ and\ \bibinfo {author} {\bibfnamefont {S.~F.}\ \bibnamefont
  {Huelga}},\ }\href@noop {} {\emph {\bibinfo {title} {Open Quantum Systems}}}\
  (\bibinfo  {publisher} {Springer},\ \bibinfo {address} {Heidelberg},\
  \bibinfo {year} {2012})\BibitemShut {NoStop}%
\bibitem [{\citenamefont {Aihara}\ \emph {et~al.}(1990)\citenamefont {Aihara},
  \citenamefont {Sevian},\ and\ \citenamefont {Skinner}}]{Aihara_PRA90}%
  \BibitemOpen
  \bibfield  {author} {\bibinfo {author} {\bibfnamefont {M.}~\bibnamefont
  {Aihara}}, \bibinfo {author} {\bibfnamefont {H.~M.}\ \bibnamefont {Sevian}},\
  and\ \bibinfo {author} {\bibfnamefont {J.~L.}\ \bibnamefont {Skinner}},\
  }\bibfield  {title} {\bibinfo {title} {Non-markovian relaxation of a
  spin-$\frac{1}{2}$ particle in a fluctuating transverse field: Cumulant
  expansion and stochastic simulation results},\ }\href@noop {} {\bibfield
  {journal} {\bibinfo  {journal} {Phys. Rev. A}\ }\textbf {\bibinfo {volume}
  {41}},\ \bibinfo {pages} {6596} (\bibinfo {year} {1990})}\BibitemShut
  {NoStop}%
\bibitem [{\citenamefont {H.~Risken}\ and\ \citenamefont
  {Vogel}(1990)}]{Risken_PRA90}%
  \BibitemOpen
  \bibfield  {author} {\bibinfo {author} {\bibfnamefont {L.~S.}\ \bibnamefont
  {H.~Risken}}\ and\ \bibinfo {author} {\bibfnamefont {K.}~\bibnamefont
  {Vogel}},\ }\bibfield  {title} {\bibinfo {title} {Relaxation of a
  spin-$\frac{1}{2}$ particle driven by transverse colored gaussian noise: Time
  dependence and eigenvalues},\ }\href@noop {} {\bibfield  {journal} {\bibinfo
  {journal} {Phys. Rev. A}\ }\textbf {\bibinfo {volume} {42}},\ \bibinfo
  {pages} {4562} (\bibinfo {year} {1990})}\BibitemShut {NoStop}%
\bibitem [{\citenamefont {Falci}\ \emph {et~al.}(2005)\citenamefont {Falci},
  \citenamefont {D'Arrigo}, \citenamefont {Mastellone},\ and\ \citenamefont
  {Paladino}}]{Falci_PRL05}%
  \BibitemOpen
  \bibfield  {author} {\bibinfo {author} {\bibfnamefont {G.}~\bibnamefont
  {Falci}}, \bibinfo {author} {\bibfnamefont {A.}~\bibnamefont {D'Arrigo}},
  \bibinfo {author} {\bibfnamefont {A.}~\bibnamefont {Mastellone}},\ and\
  \bibinfo {author} {\bibfnamefont {E.}~\bibnamefont {Paladino}},\ }\bibfield
  {title} {\bibinfo {title} {Initial decoherence in solid state qubits},\
  }\href {https://doi.org/10.1103/PhysRevLett.94.167002} {\bibfield  {journal}
  {\bibinfo  {journal} {Phys.\ Rev.\ Lett.}\ }\textbf {\bibinfo {volume}
  {94}},\ \bibinfo {pages} {167002} (\bibinfo {year} {2005})}\BibitemShut
  {NoStop}%
\bibitem [{\citenamefont {Barnes}\ \emph {et~al.}(2016)\citenamefont {Barnes},
  \citenamefont {Rudner}, \citenamefont {Martins}, \citenamefont {Malinowski},
  \citenamefont {Marcus},\ and\ \citenamefont {Kuemmeth}}]{Barnes_PRB16}%
  \BibitemOpen
  \bibfield  {author} {\bibinfo {author} {\bibfnamefont {E.}~\bibnamefont
  {Barnes}}, \bibinfo {author} {\bibfnamefont {M.~S.}\ \bibnamefont {Rudner}},
  \bibinfo {author} {\bibfnamefont {F.}~\bibnamefont {Martins}}, \bibinfo
  {author} {\bibfnamefont {F.~K.}\ \bibnamefont {Malinowski}}, \bibinfo
  {author} {\bibfnamefont {C.~M.}\ \bibnamefont {Marcus}},\ and\ \bibinfo
  {author} {\bibfnamefont {F.}~\bibnamefont {Kuemmeth}},\ }\bibfield  {title}
  {\bibinfo {title} {Filter function formalism beyond pure dephasing and
  non-markovian noise in singlet-triplet qubits},\ }\href
  {https://doi.org/10.1103/PhysRevB.93.121407} {\bibfield  {journal} {\bibinfo
  {journal} {Phys. Rev. B}\ }\textbf {\bibinfo {volume} {93}},\ \bibinfo
  {pages} {121407} (\bibinfo {year} {2016})}\BibitemShut {NoStop}%
\bibitem [{\citenamefont {Viola}\ \emph {et~al.}(1999)\citenamefont {Viola},
  \citenamefont {Knill},\ and\ \citenamefont {Lloyd}}]{Viola_PRL99}%
  \BibitemOpen
  \bibfield  {author} {\bibinfo {author} {\bibfnamefont {L.}~\bibnamefont
  {Viola}}, \bibinfo {author} {\bibfnamefont {E.}~\bibnamefont {Knill}},\ and\
  \bibinfo {author} {\bibfnamefont {S.}~\bibnamefont {Lloyd}},\ }\bibfield
  {title} {\bibinfo {title} {Dynamical decoupling of open quantum systems},\
  }\href {https://doi.org/10.1103/PhysRevLett.82.2417} {\bibfield  {journal}
  {\bibinfo  {journal} {Phys. Rev. Lett.}\ }\textbf {\bibinfo {volume} {82}},\
  \bibinfo {pages} {2417} (\bibinfo {year} {1999})}\BibitemShut {NoStop}%
\bibitem [{\citenamefont {Kofman}\ and\ \citenamefont
  {Kurizki}(2004)}]{Kofman_PRL04}%
  \BibitemOpen
  \bibfield  {author} {\bibinfo {author} {\bibfnamefont {A.~G.}\ \bibnamefont
  {Kofman}}\ and\ \bibinfo {author} {\bibfnamefont {G.}~\bibnamefont
  {Kurizki}},\ }\bibfield  {title} {\bibinfo {title} {Unified theory of
  dynamically suppressed qubit decoherence in thermal baths},\ }\href
  {https://doi.org/10.1103/PhysRevLett.93.130406} {\bibfield  {journal}
  {\bibinfo  {journal} {Phys. Rev. Lett.}\ }\textbf {\bibinfo {volume} {93}},\
  \bibinfo {pages} {130406} (\bibinfo {year} {2004})}\BibitemShut {NoStop}%
\bibitem [{\citenamefont {Biercuk}\ \emph {et~al.}(2011)\citenamefont
  {Biercuk}, \citenamefont {Doherty},\ and\ \citenamefont
  {Uys}}]{Biercuk_JPB11}%
  \BibitemOpen
  \bibfield  {author} {\bibinfo {author} {\bibfnamefont {M.~J.}\ \bibnamefont
  {Biercuk}}, \bibinfo {author} {\bibfnamefont {A.~C.}\ \bibnamefont
  {Doherty}},\ and\ \bibinfo {author} {\bibfnamefont {H.}~\bibnamefont {Uys}},\
  }\bibfield  {title} {\bibinfo {title} {Dynamical decoupling sequence
  construction as a filter-design problem},\ }\href
  {https://doi.org/10.1088/0953-4075/44/15/154002} {\bibfield  {journal}
  {\bibinfo  {journal} {J.~Phys.~B: At.~Mol.~Opt.~Phys.}\ }\textbf {\bibinfo
  {volume} {44}},\ \bibinfo {pages} {154002} (\bibinfo {year}
  {2011})}\BibitemShut {NoStop}%
\bibitem [{\citenamefont {Lidar}(2014)}]{Lidar_ACP14}%
  \BibitemOpen
  \bibfield  {author} {\bibinfo {author} {\bibfnamefont {D.~A.}\ \bibnamefont
  {Lidar}},\ }\bibfield  {title} {\bibinfo {title} {Review of decoherence-free
  subspaces, noiseless subsystems, and dynamical decoupling},\ }\href
  {https://doi.org/10.1002/9781118742631.ch11} {\bibfield  {journal} {\bibinfo
  {journal} {Adv.~Chem.~Phys.}\ }\textbf {\bibinfo {volume} {154}},\ \bibinfo
  {pages} {295} (\bibinfo {year} {2014})}\BibitemShut {NoStop}%
\bibitem [{\citenamefont {Suter}\ and\ \citenamefont
  {\'Alvarez}(2016)}]{Suter_RMP16}%
  \BibitemOpen
  \bibfield  {author} {\bibinfo {author} {\bibfnamefont {D.}~\bibnamefont
  {Suter}}\ and\ \bibinfo {author} {\bibfnamefont {G.~A.}\ \bibnamefont
  {\'Alvarez}},\ }\bibfield  {title} {\bibinfo {title} {Colloquium: Protecting
  quantum information against environmental noise},\ }\href
  {https://doi.org/10.1103/RevModPhys.88.041001} {\bibfield  {journal}
  {\bibinfo  {journal} {Rev. Mod. Phys.}\ }\textbf {\bibinfo {volume} {88}},\
  \bibinfo {pages} {041001} (\bibinfo {year} {2016})}\BibitemShut {NoStop}%
\bibitem [{\citenamefont {Degen}\ \emph {et~al.}(2017)\citenamefont {Degen},
  \citenamefont {Reinhard},\ and\ \citenamefont {Cappellaro}}]{Degen_RMP17}%
  \BibitemOpen
  \bibfield  {author} {\bibinfo {author} {\bibfnamefont {C.~L.}\ \bibnamefont
  {Degen}}, \bibinfo {author} {\bibfnamefont {F.}~\bibnamefont {Reinhard}},\
  and\ \bibinfo {author} {\bibfnamefont {P.}~\bibnamefont {Cappellaro}},\
  }\bibfield  {title} {\bibinfo {title} {Quantum sensing},\ }\href
  {https://doi.org/10.1103/RevModPhys.89.035002} {\bibfield  {journal}
  {\bibinfo  {journal} {Rev. Mod. Phys.}\ }\textbf {\bibinfo {volume} {89}},\
  \bibinfo {pages} {035002} (\bibinfo {year} {2017})}\BibitemShut {NoStop}%
\bibitem [{\citenamefont {Petta}\ \emph {et~al.}(2005)\citenamefont {Petta},
  \citenamefont {Johnson}, \citenamefont {Taylor}, \citenamefont {Laird},
  \citenamefont {Yacoby}, \citenamefont {Lukin}, \citenamefont {Marcus},
  \citenamefont {Hanson},\ and\ \citenamefont {Gossard}}]{Petta_Science05}%
  \BibitemOpen
  \bibfield  {author} {\bibinfo {author} {\bibfnamefont {J.~R.}\ \bibnamefont
  {Petta}}, \bibinfo {author} {\bibfnamefont {A.~C.}\ \bibnamefont {Johnson}},
  \bibinfo {author} {\bibfnamefont {J.~M.}\ \bibnamefont {Taylor}}, \bibinfo
  {author} {\bibfnamefont {E.~A.}\ \bibnamefont {Laird}}, \bibinfo {author}
  {\bibfnamefont {A.}~\bibnamefont {Yacoby}}, \bibinfo {author} {\bibfnamefont
  {M.~D.}\ \bibnamefont {Lukin}}, \bibinfo {author} {\bibfnamefont {C.~M.}\
  \bibnamefont {Marcus}}, \bibinfo {author} {\bibfnamefont {M.~P.}\
  \bibnamefont {Hanson}},\ and\ \bibinfo {author} {\bibfnamefont {A.~C.}\
  \bibnamefont {Gossard}},\ }\bibfield  {title} {\bibinfo {title} {{Coherent
  Manipulation of Coupled Electron Spins in Semiconductor Quantum Dots}},\
  }\href {https://doi.org/10.1126/science.1116955} {\bibfield  {journal}
  {\bibinfo  {journal} {Science}\ }\textbf {\bibinfo {volume} {309}},\ \bibinfo
  {pages} {2180} (\bibinfo {year} {2005})}\BibitemShut {NoStop}%
\bibitem [{\citenamefont {Taylor}\ \emph {et~al.}(2005)\citenamefont {Taylor},
  \citenamefont {Engel}, \citenamefont {D\"{u}r}, \citenamefont {Yacoby},
  \citenamefont {Marcus}, \citenamefont {Zoller},\ and\ \citenamefont
  {Lukin}}]{Taylor_NatPhys05}%
  \BibitemOpen
  \bibfield  {author} {\bibinfo {author} {\bibfnamefont {J.~M.}\ \bibnamefont
  {Taylor}}, \bibinfo {author} {\bibfnamefont {H.}~\bibnamefont {Engel}},
  \bibinfo {author} {\bibfnamefont {W.}~\bibnamefont {D\"{u}r}}, \bibinfo
  {author} {\bibfnamefont {A.}~\bibnamefont {Yacoby}}, \bibinfo {author}
  {\bibfnamefont {C.~M.}\ \bibnamefont {Marcus}}, \bibinfo {author}
  {\bibfnamefont {P.}~\bibnamefont {Zoller}},\ and\ \bibinfo {author}
  {\bibfnamefont {M.~D.}\ \bibnamefont {Lukin}},\ }\bibfield  {title} {\bibinfo
  {title} {Fault-tolerant architecture for quantum computation using
  electrically controlled semiconductor spins},\ }\href
  {https://doi.org/10.1038/nphys174} {\bibfield  {journal} {\bibinfo  {journal}
  {Nat.~Phys.}\ }\textbf {\bibinfo {volume} {1}},\ \bibinfo {pages} {177}
  (\bibinfo {year} {2005})}\BibitemShut {NoStop}%
\bibitem [{\citenamefont {Taylor}\ \emph {et~al.}(2007)\citenamefont {Taylor},
  \citenamefont {Petta}, \citenamefont {Johnson}, \citenamefont {Yacoby},
  \citenamefont {Marcus},\ and\ \citenamefont {Lukin}}]{Taylor_PRB07}%
  \BibitemOpen
  \bibfield  {author} {\bibinfo {author} {\bibfnamefont {J.~M.}\ \bibnamefont
  {Taylor}}, \bibinfo {author} {\bibfnamefont {J.~R.}\ \bibnamefont {Petta}},
  \bibinfo {author} {\bibfnamefont {A.~C.}\ \bibnamefont {Johnson}}, \bibinfo
  {author} {\bibfnamefont {A.}~\bibnamefont {Yacoby}}, \bibinfo {author}
  {\bibfnamefont {C.~M.}\ \bibnamefont {Marcus}},\ and\ \bibinfo {author}
  {\bibfnamefont {M.~D.}\ \bibnamefont {Lukin}},\ }\bibfield  {title} {\bibinfo
  {title} {Relaxation, dephasing, and quantum control of electron spins in
  double quantum dots},\ }\href {https://doi.org/10.1103/PhysRevB.76.035315}
  {\bibfield  {journal} {\bibinfo  {journal} {Phys.\ Rev.\ B}\ }\textbf
  {\bibinfo {volume} {76}},\ \bibinfo {pages} {035315} (\bibinfo {year}
  {2007})}\BibitemShut {NoStop}%
\bibitem [{\citenamefont {Medford}\ \emph
  {et~al.}(2013{\natexlab{a}})\citenamefont {Medford}, \citenamefont {Beil},
  \citenamefont {Taylor}, \citenamefont {Rashba}, \citenamefont {Lu},
  \citenamefont {Gossard},\ and\ \citenamefont {Marcus}}]{Medford_PRL13}%
  \BibitemOpen
  \bibfield  {author} {\bibinfo {author} {\bibfnamefont {J.}~\bibnamefont
  {Medford}}, \bibinfo {author} {\bibfnamefont {J.}~\bibnamefont {Beil}},
  \bibinfo {author} {\bibfnamefont {J.~M.}\ \bibnamefont {Taylor}}, \bibinfo
  {author} {\bibfnamefont {E.~I.}\ \bibnamefont {Rashba}}, \bibinfo {author}
  {\bibfnamefont {H.}~\bibnamefont {Lu}}, \bibinfo {author} {\bibfnamefont
  {A.~C.}\ \bibnamefont {Gossard}},\ and\ \bibinfo {author} {\bibfnamefont
  {C.~M.}\ \bibnamefont {Marcus}},\ }\bibfield  {title} {\bibinfo {title}
  {Quantum-dot-based resonant exchange qubit},\ }\href
  {https://doi.org/10.1103/PhysRevLett.111.050501} {\bibfield  {journal}
  {\bibinfo  {journal} {Phys.\ Rev.\ Lett.}\ }\textbf {\bibinfo {volume}
  {111}},\ \bibinfo {pages} {050501} (\bibinfo {year}
  {2013}{\natexlab{a}})}\BibitemShut {NoStop}%
\bibitem [{\citenamefont {Medford}\ \emph
  {et~al.}(2013{\natexlab{b}})\citenamefont {Medford}, \citenamefont {Beil},
  \citenamefont {Taylor}, \citenamefont {Bartlett}, \citenamefont {Doherty},
  \citenamefont {Rashba}, \citenamefont {DiVincenzo}, \citenamefont {Lu},
  \citenamefont {Gossard},\ and\ \citenamefont {Marcus}}]{Medford_NN13}%
  \BibitemOpen
  \bibfield  {author} {\bibinfo {author} {\bibfnamefont {J.}~\bibnamefont
  {Medford}}, \bibinfo {author} {\bibfnamefont {J.}~\bibnamefont {Beil}},
  \bibinfo {author} {\bibfnamefont {J.~M.}\ \bibnamefont {Taylor}}, \bibinfo
  {author} {\bibfnamefont {S.~D.}\ \bibnamefont {Bartlett}}, \bibinfo {author}
  {\bibfnamefont {A.~C.}\ \bibnamefont {Doherty}}, \bibinfo {author}
  {\bibfnamefont {E.~I.}\ \bibnamefont {Rashba}}, \bibinfo {author}
  {\bibfnamefont {D.~P.}\ \bibnamefont {DiVincenzo}}, \bibinfo {author}
  {\bibfnamefont {H.}~\bibnamefont {Lu}}, \bibinfo {author} {\bibfnamefont
  {A.~C.}\ \bibnamefont {Gossard}},\ and\ \bibinfo {author} {\bibfnamefont
  {C.~M.}\ \bibnamefont {Marcus}},\ }\bibfield  {title} {\bibinfo {title}
  {Self-consistent measurement and state tomography of an exchange-only spin
  qubit},\ }\href {https://doi.org/10.1038/nnano.2013.168} {\bibfield
  {journal} {\bibinfo  {journal} {Nature Nanotechnology}\ }\textbf {\bibinfo
  {volume} {8}},\ \bibinfo {pages} {654} (\bibinfo {year}
  {2013}{\natexlab{b}})}\BibitemShut {NoStop}%
\bibitem [{\citenamefont {Russ}\ \emph {et~al.}(2016)\citenamefont {Russ},
  \citenamefont {Ginzel},\ and\ \citenamefont {Burkard}}]{Russ_PRB16}%
  \BibitemOpen
  \bibfield  {author} {\bibinfo {author} {\bibfnamefont {M.}~\bibnamefont
  {Russ}}, \bibinfo {author} {\bibfnamefont {F.}~\bibnamefont {Ginzel}},\ and\
  \bibinfo {author} {\bibfnamefont {G.}~\bibnamefont {Burkard}},\ }\bibfield
  {title} {\bibinfo {title} {Coupling of three-spin qubits to their electric
  environment},\ }\href {https://doi.org/10.1103/PhysRevB.94.165411} {\bibfield
   {journal} {\bibinfo  {journal} {Phys. Rev. B}\ }\textbf {\bibinfo {volume}
  {94}},\ \bibinfo {pages} {165411} (\bibinfo {year} {2016})}\BibitemShut
  {NoStop}%
\bibitem [{\citenamefont {Sala}\ and\ \citenamefont
  {Danon}(2017)}]{Sala_PRB17}%
  \BibitemOpen
  \bibfield  {author} {\bibinfo {author} {\bibfnamefont {A.}~\bibnamefont
  {Sala}}\ and\ \bibinfo {author} {\bibfnamefont {J.}~\bibnamefont {Danon}},\
  }\bibfield  {title} {\bibinfo {title} {Exchange-only singlet-only spin
  qubit},\ }\href {https://doi.org/10.1103/PhysRevB.95.241303} {\bibfield
  {journal} {\bibinfo  {journal} {Phys. Rev. B}\ }\textbf {\bibinfo {volume}
  {95}},\ \bibinfo {pages} {241303} (\bibinfo {year} {2017})}\BibitemShut
  {NoStop}%
\bibitem [{\citenamefont {Russ}\ \emph {et~al.}(2018)\citenamefont {Russ},
  \citenamefont {Petta},\ and\ \citenamefont {Burkard}}]{Russ_PRL18}%
  \BibitemOpen
  \bibfield  {author} {\bibinfo {author} {\bibfnamefont {M.}~\bibnamefont
  {Russ}}, \bibinfo {author} {\bibfnamefont {J.~R.}\ \bibnamefont {Petta}},\
  and\ \bibinfo {author} {\bibfnamefont {G.}~\bibnamefont {Burkard}},\
  }\bibfield  {title} {\bibinfo {title} {Quadrupolar exchange-only spin
  qubit},\ }\href {https://doi.org/10.1103/PhysRevLett.121.177701} {\bibfield
  {journal} {\bibinfo  {journal} {Phys. Rev. Lett.}\ }\textbf {\bibinfo
  {volume} {121}},\ \bibinfo {pages} {177701} (\bibinfo {year}
  {2018})}\BibitemShut {NoStop}%
\bibitem [{\citenamefont {Bylander}\ \emph {et~al.}(2011)\citenamefont
  {Bylander}, \citenamefont {Gustavsson}, \citenamefont {Yan}, \citenamefont
  {Yoshihara}, \citenamefont {Harrabi}, \citenamefont {Fitch}, \citenamefont
  {Cory}, \citenamefont {Nakamura}, \citenamefont {Tsai},\ and\ \citenamefont
  {Oliver}}]{Bylander_NatPhys11}%
  \BibitemOpen
  \bibfield  {author} {\bibinfo {author} {\bibfnamefont {J.}~\bibnamefont
  {Bylander}}, \bibinfo {author} {\bibfnamefont {S.}~\bibnamefont
  {Gustavsson}}, \bibinfo {author} {\bibfnamefont {F.}~\bibnamefont {Yan}},
  \bibinfo {author} {\bibfnamefont {F.}~\bibnamefont {Yoshihara}}, \bibinfo
  {author} {\bibfnamefont {K.}~\bibnamefont {Harrabi}}, \bibinfo {author}
  {\bibfnamefont {G.}~\bibnamefont {Fitch}}, \bibinfo {author} {\bibfnamefont
  {D.~G.}\ \bibnamefont {Cory}}, \bibinfo {author} {\bibfnamefont
  {Y.}~\bibnamefont {Nakamura}}, \bibinfo {author} {\bibfnamefont {J.-S.}\
  \bibnamefont {Tsai}},\ and\ \bibinfo {author} {\bibfnamefont {W.~D.}\
  \bibnamefont {Oliver}},\ }\bibfield  {title} {\bibinfo {title} {Noise
  spectroscopy through dynamical decoupling with a superconducting flux
  qubit},\ }\href {https://doi.org/10.1038/nphys1994} {\bibfield  {journal}
  {\bibinfo  {journal} {Nature Physics}\ }\textbf {\bibinfo {volume} {7}},\
  \bibinfo {pages} {565} (\bibinfo {year} {2011})}\BibitemShut {NoStop}%
\bibitem [{\citenamefont {Yan}\ \emph {et~al.}(2016)\citenamefont {Yan},
  \citenamefont {Gustavsson}, \citenamefont {Kamal}, \citenamefont {Birenbaum},
  \citenamefont {Sears}, \citenamefont {Hover}, \citenamefont {Gudmundsen},
  \citenamefont {Rosenberg}, \citenamefont {Samach}, \citenamefont {Weber},
  \citenamefont {Yoder}, \citenamefont {Orlando}, \citenamefont {Clarke},
  \citenamefont {Kerman},\ and\ \citenamefont {Oliver}}]{Yan_NatComm16}%
  \BibitemOpen
  \bibfield  {author} {\bibinfo {author} {\bibfnamefont {F.}~\bibnamefont
  {Yan}}, \bibinfo {author} {\bibfnamefont {S.}~\bibnamefont {Gustavsson}},
  \bibinfo {author} {\bibfnamefont {A.}~\bibnamefont {Kamal}}, \bibinfo
  {author} {\bibfnamefont {J.}~\bibnamefont {Birenbaum}}, \bibinfo {author}
  {\bibfnamefont {A.~P.}\ \bibnamefont {Sears}}, \bibinfo {author}
  {\bibfnamefont {D.}~\bibnamefont {Hover}}, \bibinfo {author} {\bibfnamefont
  {T.~J.}\ \bibnamefont {Gudmundsen}}, \bibinfo {author} {\bibfnamefont
  {D.}~\bibnamefont {Rosenberg}}, \bibinfo {author} {\bibfnamefont
  {G.}~\bibnamefont {Samach}}, \bibinfo {author} {\bibfnamefont
  {S.}~\bibnamefont {Weber}}, \bibinfo {author} {\bibfnamefont {J.~L.}\
  \bibnamefont {Yoder}}, \bibinfo {author} {\bibfnamefont {T.~P.}\ \bibnamefont
  {Orlando}}, \bibinfo {author} {\bibfnamefont {J.}~\bibnamefont {Clarke}},
  \bibinfo {author} {\bibfnamefont {A.~J.}\ \bibnamefont {Kerman}},\ and\
  \bibinfo {author} {\bibfnamefont {W.~D.}\ \bibnamefont {Oliver}},\ }\bibfield
   {title} {\bibinfo {title} {The flux qubit revisited to enhance coherence and
  reproducibility},\ }\href {https://doi.org/10.1038/ncomms12964} {\bibfield
  {journal} {\bibinfo  {journal} {Nature Communications}\ }\textbf {\bibinfo
  {volume} {7}},\ \bibinfo {pages} {12964} (\bibinfo {year}
  {2016})}\BibitemShut {NoStop}%
\bibitem [{\citenamefont {Krantz}\ \emph {et~al.}(2019)\citenamefont {Krantz},
  \citenamefont {Kjaergaard}, \citenamefont {Yan}, \citenamefont {Orlando},
  \citenamefont {Gustavsson},\ and\ \citenamefont {Oliver}}]{Krantz_APR19}%
  \BibitemOpen
  \bibfield  {author} {\bibinfo {author} {\bibfnamefont {P.}~\bibnamefont
  {Krantz}}, \bibinfo {author} {\bibfnamefont {M.}~\bibnamefont {Kjaergaard}},
  \bibinfo {author} {\bibfnamefont {F.}~\bibnamefont {Yan}}, \bibinfo {author}
  {\bibfnamefont {T.~P.}\ \bibnamefont {Orlando}}, \bibinfo {author}
  {\bibfnamefont {S.}~\bibnamefont {Gustavsson}},\ and\ \bibinfo {author}
  {\bibfnamefont {W.~D.}\ \bibnamefont {Oliver}},\ }\bibfield  {title}
  {\bibinfo {title} {A quantum engineer’s guide to superconducting qubits},\
  }\href {https://doi.org/10.1063/1.5089550} {\bibfield  {journal} {\bibinfo
  {journal} {Applied Physics Reviews}\ }\textbf {\bibinfo {volume} {6}},\
  \bibinfo {pages} {021318} (\bibinfo {year} {2019})}\BibitemShut {NoStop}%
\bibitem [{\citenamefont {Nakamura}\ \emph {et~al.}(2002)\citenamefont
  {Nakamura}, \citenamefont {Pashkin}, \citenamefont {Yamamoto},\ and\
  \citenamefont {Tsai}}]{Nakamura_PRL02}%
  \BibitemOpen
  \bibfield  {author} {\bibinfo {author} {\bibfnamefont {Y.}~\bibnamefont
  {Nakamura}}, \bibinfo {author} {\bibfnamefont {Y.~A.}\ \bibnamefont
  {Pashkin}}, \bibinfo {author} {\bibfnamefont {T.}~\bibnamefont {Yamamoto}},\
  and\ \bibinfo {author} {\bibfnamefont {J.~S.}\ \bibnamefont {Tsai}},\
  }\bibfield  {title} {\bibinfo {title} {Charge echo in a cooper-pair box},\
  }\href@noop {} {\bibfield  {journal} {\bibinfo  {journal} {Phys.\ Rev.\
  Lett.}\ }\textbf {\bibinfo {volume} {88}},\ \bibinfo {pages} {047901}
  (\bibinfo {year} {2002})}\BibitemShut {NoStop}%
\bibitem [{\citenamefont {Astafiev}\ \emph {et~al.}(2004)\citenamefont
  {Astafiev}, \citenamefont {Pashkin}, \citenamefont {Nakamura}, \citenamefont
  {Yamamoto},\ and\ \citenamefont {Tsai}}]{Astafiev_PRL04}%
  \BibitemOpen
  \bibfield  {author} {\bibinfo {author} {\bibfnamefont {O.}~\bibnamefont
  {Astafiev}}, \bibinfo {author} {\bibfnamefont {Y.~A.}\ \bibnamefont
  {Pashkin}}, \bibinfo {author} {\bibfnamefont {Y.}~\bibnamefont {Nakamura}},
  \bibinfo {author} {\bibfnamefont {T.}~\bibnamefont {Yamamoto}},\ and\
  \bibinfo {author} {\bibfnamefont {J.~S.}\ \bibnamefont {Tsai}},\ }\bibfield
  {title} {\bibinfo {title} {Quantum noise in the josephson charge qubit},\
  }\href {https://doi.org/10.1103/PhysRevLett.93.267007} {\bibfield  {journal}
  {\bibinfo  {journal} {Phys.\ Rev.\ Lett.}\ }\textbf {\bibinfo {volume}
  {93}},\ \bibinfo {pages} {267007} (\bibinfo {year} {2004})}\BibitemShut
  {NoStop}%
\bibitem [{\citenamefont {Jung}\ \emph {et~al.}(2004)\citenamefont {Jung},
  \citenamefont {Fujisawa}, \citenamefont {Hirayama},\ and\ \citenamefont
  {Jeong}}]{Jung_APL04}%
  \BibitemOpen
  \bibfield  {author} {\bibinfo {author} {\bibfnamefont {S.~W.}\ \bibnamefont
  {Jung}}, \bibinfo {author} {\bibfnamefont {T.}~\bibnamefont {Fujisawa}},
  \bibinfo {author} {\bibfnamefont {Y.}~\bibnamefont {Hirayama}},\ and\
  \bibinfo {author} {\bibfnamefont {Y.~H.}\ \bibnamefont {Jeong}},\ }\bibfield
  {title} {\bibinfo {title} {Background charge fluctuation in a gaas quantum
  dot device},\ }\href {https://doi.org/10.1063/1.1777802} {\bibfield
  {journal} {\bibinfo  {journal} {Appl.~Phys.~Lett.}\ }\textbf {\bibinfo
  {volume} {85}},\ \bibinfo {pages} {768} (\bibinfo {year} {2004})}\BibitemShut
  {NoStop}%
\bibitem [{\citenamefont {Gustafsson}\ \emph {et~al.}(2013)\citenamefont
  {Gustafsson}, \citenamefont {Pourkabirian}, \citenamefont {Johansson},
  \citenamefont {Clarke},\ and\ \citenamefont {Delsing}}]{Gustafsson_PRB13}%
  \BibitemOpen
  \bibfield  {author} {\bibinfo {author} {\bibfnamefont {M.~V.}\ \bibnamefont
  {Gustafsson}}, \bibinfo {author} {\bibfnamefont {A.}~\bibnamefont
  {Pourkabirian}}, \bibinfo {author} {\bibfnamefont {G.}~\bibnamefont
  {Johansson}}, \bibinfo {author} {\bibfnamefont {J.}~\bibnamefont {Clarke}},\
  and\ \bibinfo {author} {\bibfnamefont {P.}~\bibnamefont {Delsing}},\
  }\bibfield  {title} {\bibinfo {title} {Thermal properties of charge noise
  sources},\ }\href {https://doi.org/10.1103/PhysRevB.88.245410} {\bibfield
  {journal} {\bibinfo  {journal} {Phys. Rev. B}\ }\textbf {\bibinfo {volume}
  {88}},\ \bibinfo {pages} {245410} (\bibinfo {year} {2013})}\BibitemShut
  {NoStop}%
\bibitem [{\citenamefont {Blake M.~Freeman}\ and\ \citenamefont
  {Jiang}(2004)}]{Freeman_APL16}%
  \BibitemOpen
  \bibfield  {author} {\bibinfo {author} {\bibfnamefont {J.~S.~S.}\
  \bibnamefont {Blake M.~Freeman}}\ and\ \bibinfo {author} {\bibfnamefont
  {H.~W.}\ \bibnamefont {Jiang}},\ }\bibfield  {title} {\bibinfo {title}
  {Comparison of low frequency charge noise in identically patterned
  si/sio$_{2}$ and si/sige quantum dots},\ }\href
  {https://doi.org/10.1063/1.4954700} {\bibfield  {journal} {\bibinfo
  {journal} {Appl.~Phys.~Lett.}\ }\textbf {\bibinfo {volume} {108}},\ \bibinfo
  {pages} {253108} (\bibinfo {year} {2004})}\BibitemShut {NoStop}%
\bibitem [{\citenamefont {Dial}\ \emph {et~al.}(2013)\citenamefont {Dial},
  \citenamefont {Shulman}, \citenamefont {Harvey}, \citenamefont {Bluhm},
  \citenamefont {Umansky},\ and\ \citenamefont {Yacoby}}]{Dial_PRL13}%
  \BibitemOpen
  \bibfield  {author} {\bibinfo {author} {\bibfnamefont {O.~E.}\ \bibnamefont
  {Dial}}, \bibinfo {author} {\bibfnamefont {M.~D.}\ \bibnamefont {Shulman}},
  \bibinfo {author} {\bibfnamefont {S.~P.}\ \bibnamefont {Harvey}}, \bibinfo
  {author} {\bibfnamefont {H.}~\bibnamefont {Bluhm}}, \bibinfo {author}
  {\bibfnamefont {V.}~\bibnamefont {Umansky}},\ and\ \bibinfo {author}
  {\bibfnamefont {A.}~\bibnamefont {Yacoby}},\ }\bibfield  {title} {\bibinfo
  {title} {Charge noise spectroscopy using coherent exchange oscillations in a
  singlet-triplet qubit},\ }\href
  {https://doi.org/10.1103/PhysRevLett.110.146804} {\bibfield  {journal}
  {\bibinfo  {journal} {Phys.\ Rev.\ Lett.}\ }\textbf {\bibinfo {volume}
  {110}},\ \bibinfo {pages} {146804} (\bibinfo {year} {2013})}\BibitemShut
  {NoStop}%
\bibitem [{\citenamefont {Christensen}\ \emph {et~al.}(2019)\citenamefont
  {Christensen}, \citenamefont {Wilen}, \citenamefont {Opremcak}, \citenamefont
  {Nelson}, \citenamefont {Schlenker}, \citenamefont {Zimonick}, \citenamefont
  {Faoro}, \citenamefont {Ioffe}, \citenamefont {Rosen}, \citenamefont
  {DuBois}, \citenamefont {Plourde},\ and\ \citenamefont
  {McDermott}}]{Christensen_PRB19}%
  \BibitemOpen
  \bibfield  {author} {\bibinfo {author} {\bibfnamefont {B.~G.}\ \bibnamefont
  {Christensen}}, \bibinfo {author} {\bibfnamefont {C.~D.}\ \bibnamefont
  {Wilen}}, \bibinfo {author} {\bibfnamefont {A.}~\bibnamefont {Opremcak}},
  \bibinfo {author} {\bibfnamefont {J.}~\bibnamefont {Nelson}}, \bibinfo
  {author} {\bibfnamefont {F.}~\bibnamefont {Schlenker}}, \bibinfo {author}
  {\bibfnamefont {C.~H.}\ \bibnamefont {Zimonick}}, \bibinfo {author}
  {\bibfnamefont {L.}~\bibnamefont {Faoro}}, \bibinfo {author} {\bibfnamefont
  {L.~B.}\ \bibnamefont {Ioffe}}, \bibinfo {author} {\bibfnamefont {Y.~J.}\
  \bibnamefont {Rosen}}, \bibinfo {author} {\bibfnamefont {J.~L.}\ \bibnamefont
  {DuBois}}, \bibinfo {author} {\bibfnamefont {B.~L.~T.}\ \bibnamefont
  {Plourde}},\ and\ \bibinfo {author} {\bibfnamefont {R.}~\bibnamefont
  {McDermott}},\ }\bibfield  {title} {\bibinfo {title} {Anomalous charge noise
  in superconducting qubits},\ }\href
  {https://doi.org/10.1103/PhysRevB.100.140503} {\bibfield  {journal} {\bibinfo
   {journal} {Phys. Rev. B}\ }\textbf {\bibinfo {volume} {100}},\ \bibinfo
  {pages} {140503} (\bibinfo {year} {2019})}\BibitemShut {NoStop}%
\bibitem [{\citenamefont {Struck}\ \emph {et~al.}(2020)\citenamefont {Struck},
  \citenamefont {Hollmann}, \citenamefont {Schauer}, \citenamefont {Fedorets},
  \citenamefont {Schmidbauer}, \citenamefont {Sawano}, \citenamefont {Riemann},
  \citenamefont {Abrosimov}, \citenamefont {Cywi{\'n}ski}, \citenamefont
  {Bougeard},\ and\ \citenamefont {Schreiber}}]{Struck_npjqi20}%
  \BibitemOpen
  \bibfield  {author} {\bibinfo {author} {\bibfnamefont {T.}~\bibnamefont
  {Struck}}, \bibinfo {author} {\bibfnamefont {A.}~\bibnamefont {Hollmann}},
  \bibinfo {author} {\bibfnamefont {F.}~\bibnamefont {Schauer}}, \bibinfo
  {author} {\bibfnamefont {O.}~\bibnamefont {Fedorets}}, \bibinfo {author}
  {\bibfnamefont {A.}~\bibnamefont {Schmidbauer}}, \bibinfo {author}
  {\bibfnamefont {K.}~\bibnamefont {Sawano}}, \bibinfo {author} {\bibfnamefont
  {H.}~\bibnamefont {Riemann}}, \bibinfo {author} {\bibfnamefont {N.~V.}\
  \bibnamefont {Abrosimov}}, \bibinfo {author} {\bibfnamefont
  {{\L}.}~\bibnamefont {Cywi{\'n}ski}}, \bibinfo {author} {\bibfnamefont
  {D.}~\bibnamefont {Bougeard}},\ and\ \bibinfo {author} {\bibfnamefont
  {L.~R.}\ \bibnamefont {Schreiber}},\ }\bibfield  {title} {\bibinfo {title}
  {Low-frequency spin qubit detuning noise in highly purified $^{28}$si/sige},\
  }\href@noop {} {\bibfield  {journal} {\bibinfo  {journal} {npj Quantum Inf.}\
  }\textbf {\bibinfo {volume} {6}},\ \bibinfo {pages} {40} (\bibinfo {year}
  {2020})}\BibitemShut {NoStop}%
\bibitem [{\citenamefont {Coish}\ and\ \citenamefont
  {Loss}(2005)}]{Coish_PRB05}%
  \BibitemOpen
  \bibfield  {author} {\bibinfo {author} {\bibfnamefont {W.~A.}\ \bibnamefont
  {Coish}}\ and\ \bibinfo {author} {\bibfnamefont {D.}~\bibnamefont {Loss}},\
  }\bibfield  {title} {\bibinfo {title} {Singlet-triplet decoherence due to
  nuclear spins in a double quantum dot},\ }\href
  {https://doi.org/10.1103/PhysRevB.72.125337} {\bibfield  {journal} {\bibinfo
  {journal} {Phys.\ Rev.\ B}\ }\textbf {\bibinfo {volume} {72}},\ \bibinfo
  {pages} {125337} (\bibinfo {year} {2005})}\BibitemShut {NoStop}%
\bibitem [{\citenamefont {Hu}\ and\ \citenamefont {{Das
  Sarma}}(2006)}]{Hu_PRL06}%
  \BibitemOpen
  \bibfield  {author} {\bibinfo {author} {\bibfnamefont {X.}~\bibnamefont
  {Hu}}\ and\ \bibinfo {author} {\bibfnamefont {S.}~\bibnamefont {{Das
  Sarma}}},\ }\bibfield  {title} {\bibinfo {title} {Charge-fluctuation-induced
  dephasing of exchange-coupled spin qubits},\ }\href
  {https://doi.org/10.1103/PhysRevLett.96.100501} {\bibfield  {journal}
  {\bibinfo  {journal} {Phys. Rev. Lett.}\ }\textbf {\bibinfo {volume} {96}},\
  \bibinfo {pages} {100501} (\bibinfo {year} {2006})}\BibitemShut {NoStop}%
\bibitem [{\citenamefont {Culcer}\ \emph {et~al.}(2009)\citenamefont {Culcer},
  \citenamefont {Hu},\ and\ \citenamefont {Sarma}}]{Culcer_APL09}%
  \BibitemOpen
  \bibfield  {author} {\bibinfo {author} {\bibfnamefont {D.}~\bibnamefont
  {Culcer}}, \bibinfo {author} {\bibfnamefont {X.}~\bibnamefont {Hu}},\ and\
  \bibinfo {author} {\bibfnamefont {S.~D.}\ \bibnamefont {Sarma}},\ }\bibfield
  {title} {\bibinfo {title} {Dephasing of si spin qubits due to charge noise},\
  }\href {https://doi.org/10.1063/1.3194778} {\bibfield  {journal} {\bibinfo
  {journal} {Applied Physics Letters}\ }\textbf {\bibinfo {volume} {95}},\
  \bibinfo {pages} {073102} (\bibinfo {year} {2009})}\BibitemShut {NoStop}%
\bibitem [{\citenamefont {Ramon}\ and\ \citenamefont {Hu}(2010)}]{Ramon_PRB10}%
  \BibitemOpen
  \bibfield  {author} {\bibinfo {author} {\bibfnamefont {G.}~\bibnamefont
  {Ramon}}\ and\ \bibinfo {author} {\bibfnamefont {X.}~\bibnamefont {Hu}},\
  }\bibfield  {title} {\bibinfo {title} {Decoherence of spin qubits due to a
  nearby charge fluctuator in gate-defined double dots},\ }\href
  {https://doi.org/10.1103/PhysRevB.81.045304} {\bibfield  {journal} {\bibinfo
  {journal} {Phys.\ Rev.\ B}\ }\textbf {\bibinfo {volume} {81}},\ \bibinfo
  {pages} {045304} (\bibinfo {year} {2010})}\BibitemShut {NoStop}%
\bibitem [{\citenamefont {Ramon}(2012)}]{Ramon_PRB12}%
  \BibitemOpen
  \bibfield  {author} {\bibinfo {author} {\bibfnamefont {G.}~\bibnamefont
  {Ramon}},\ }\bibfield  {title} {\bibinfo {title} {Dynamical decoupling of a
  singlet-triplet qubit afflicted by a charge fluctuator},\ }\href
  {https://doi.org/10.1103/PhysRevB.86.125317} {\bibfield  {journal} {\bibinfo
  {journal} {Phys.\ Rev.\ B}\ }\textbf {\bibinfo {volume} {86}},\ \bibinfo
  {pages} {125317} (\bibinfo {year} {2012})}\BibitemShut {NoStop}%
\bibitem [{\citenamefont {Brunner}\ \emph {et~al.}(2011)\citenamefont
  {Brunner}, \citenamefont {Shin}, \citenamefont {Obata}, \citenamefont
  {Pioro-Ladri\`ere}, \citenamefont {Kubo}, \citenamefont {Yoshida},
  \citenamefont {Taniyama}, \citenamefont {Tokura},\ and\ \citenamefont
  {Tarucha}}]{Brunner_PRL11}%
  \BibitemOpen
  \bibfield  {author} {\bibinfo {author} {\bibfnamefont {R.}~\bibnamefont
  {Brunner}}, \bibinfo {author} {\bibfnamefont {Y.-S.}\ \bibnamefont {Shin}},
  \bibinfo {author} {\bibfnamefont {T.}~\bibnamefont {Obata}}, \bibinfo
  {author} {\bibfnamefont {M.}~\bibnamefont {Pioro-Ladri\`ere}}, \bibinfo
  {author} {\bibfnamefont {T.}~\bibnamefont {Kubo}}, \bibinfo {author}
  {\bibfnamefont {K.}~\bibnamefont {Yoshida}}, \bibinfo {author} {\bibfnamefont
  {T.}~\bibnamefont {Taniyama}}, \bibinfo {author} {\bibfnamefont
  {Y.}~\bibnamefont {Tokura}},\ and\ \bibinfo {author} {\bibfnamefont
  {S.}~\bibnamefont {Tarucha}},\ }\bibfield  {title} {\bibinfo {title}
  {Two-qubit gate of combined single-spin rotation and interdot spin exchange
  in a double quantum dot},\ }\href
  {https://doi.org/10.1103/PhysRevLett.107.146801} {\bibfield  {journal}
  {\bibinfo  {journal} {Phys. Rev. Lett.}\ }\textbf {\bibinfo {volume} {107}},\
  \bibinfo {pages} {146801} (\bibinfo {year} {2011})}\BibitemShut {NoStop}%
\bibitem [{\citenamefont {Yoneda}\ \emph {et~al.}(2018)\citenamefont {Yoneda},
  \citenamefont {Takeda}, \citenamefont {Otsuka}, \citenamefont {Nakajima},
  \citenamefont {Delbecq}, \citenamefont {Allison}, \citenamefont {Honda},
  \citenamefont {Kodera}, \citenamefont {Oda}, \citenamefont {Hoshi},
  \citenamefont {Usami}, \citenamefont {Itoh},\ and\ \citenamefont
  {Tarucha}}]{Yoneda_NN18}%
  \BibitemOpen
  \bibfield  {author} {\bibinfo {author} {\bibfnamefont {J.}~\bibnamefont
  {Yoneda}}, \bibinfo {author} {\bibfnamefont {K.}~\bibnamefont {Takeda}},
  \bibinfo {author} {\bibfnamefont {T.}~\bibnamefont {Otsuka}}, \bibinfo
  {author} {\bibfnamefont {T.}~\bibnamefont {Nakajima}}, \bibinfo {author}
  {\bibfnamefont {M.~R.}\ \bibnamefont {Delbecq}}, \bibinfo {author}
  {\bibfnamefont {G.}~\bibnamefont {Allison}}, \bibinfo {author} {\bibfnamefont
  {T.}~\bibnamefont {Honda}}, \bibinfo {author} {\bibfnamefont
  {T.}~\bibnamefont {Kodera}}, \bibinfo {author} {\bibfnamefont
  {S.}~\bibnamefont {Oda}}, \bibinfo {author} {\bibfnamefont {Y.}~\bibnamefont
  {Hoshi}}, \bibinfo {author} {\bibfnamefont {N.}~\bibnamefont {Usami}},
  \bibinfo {author} {\bibfnamefont {K.~M.}\ \bibnamefont {Itoh}},\ and\
  \bibinfo {author} {\bibfnamefont {S.}~\bibnamefont {Tarucha}},\ }\bibfield
  {title} {\bibinfo {title} {A quantum-dot spin qubit with coherence limited by
  charge noise and fidelity higher than 99.9\%},\ }\href
  {https://doi.org/10.1038/s41565-017-0014-x} {\bibfield  {journal} {\bibinfo
  {journal} {Natute Nanotechnology}\ }\textbf {\bibinfo {volume} {13}},\
  \bibinfo {pages} {102} (\bibinfo {year} {2018})}\BibitemShut {NoStop}%
\bibitem [{\citenamefont {Zajac}\ \emph {et~al.}(2018)\citenamefont {Zajac},
  \citenamefont {Sigillito}, \citenamefont {Russ}, \citenamefont {Borjans},
  \citenamefont {Taylor}, \citenamefont {Burkard},\ and\ \citenamefont
  {Petta}}]{Zajac_Science19}%
  \BibitemOpen
  \bibfield  {author} {\bibinfo {author} {\bibfnamefont {D.~M.}\ \bibnamefont
  {Zajac}}, \bibinfo {author} {\bibfnamefont {A.~J.}\ \bibnamefont
  {Sigillito}}, \bibinfo {author} {\bibfnamefont {M.}~\bibnamefont {Russ}},
  \bibinfo {author} {\bibfnamefont {F.}~\bibnamefont {Borjans}}, \bibinfo
  {author} {\bibfnamefont {J.~M.}\ \bibnamefont {Taylor}}, \bibinfo {author}
  {\bibfnamefont {G.}~\bibnamefont {Burkard}},\ and\ \bibinfo {author}
  {\bibfnamefont {J.~R.}\ \bibnamefont {Petta}},\ }\bibfield  {title} {\bibinfo
  {title} {Resonantly driven cnot gate for electron spins},\ }\href
  {https://doi.org/10.1126/science.aao5965} {\bibfield  {journal} {\bibinfo
  {journal} {Science}\ }\textbf {\bibinfo {volume} {359}},\ \bibinfo {pages}
  {439} (\bibinfo {year} {2018})}\BibitemShut {NoStop}%
\bibitem [{\citenamefont {Foletti}\ \emph {et~al.}(2009)\citenamefont
  {Foletti}, \citenamefont {Bluhm}, \citenamefont {Mahalu}, \citenamefont
  {Umansky},\ and\ \citenamefont {Yacoby}}]{Foletti_NatPhys09}%
  \BibitemOpen
  \bibfield  {author} {\bibinfo {author} {\bibfnamefont {S.}~\bibnamefont
  {Foletti}}, \bibinfo {author} {\bibfnamefont {H.}~\bibnamefont {Bluhm}},
  \bibinfo {author} {\bibfnamefont {D.}~\bibnamefont {Mahalu}}, \bibinfo
  {author} {\bibfnamefont {V.}~\bibnamefont {Umansky}},\ and\ \bibinfo {author}
  {\bibfnamefont {A.}~\bibnamefont {Yacoby}},\ }\bibfield  {title} {\bibinfo
  {title} {Universal quantum control of two-electron spin quantum bits using
  dynamic nuclear polarization},\ }\href {https://doi.org/10.1038/NPHYS1424}
  {\bibfield  {journal} {\bibinfo  {journal} {Nat.~Phys.}\ }\textbf {\bibinfo
  {volume} {5}},\ \bibinfo {pages} {903} (\bibinfo {year} {2009})}\BibitemShut
  {NoStop}%
\bibitem [{\citenamefont {Bluhm}\ \emph
  {et~al.}(2010{\natexlab{a}})\citenamefont {Bluhm}, \citenamefont {Foletti},
  \citenamefont {Mahalu}, \citenamefont {Umansky},\ and\ \citenamefont
  {Yacoby}}]{Bluhm_PRL10}%
  \BibitemOpen
  \bibfield  {author} {\bibinfo {author} {\bibfnamefont {H.}~\bibnamefont
  {Bluhm}}, \bibinfo {author} {\bibfnamefont {S.}~\bibnamefont {Foletti}},
  \bibinfo {author} {\bibfnamefont {D.}~\bibnamefont {Mahalu}}, \bibinfo
  {author} {\bibfnamefont {V.}~\bibnamefont {Umansky}},\ and\ \bibinfo {author}
  {\bibfnamefont {A.}~\bibnamefont {Yacoby}},\ }\bibfield  {title} {\bibinfo
  {title} {Enhancing the coherence of a spin qubit by operating it as a
  feedback loop that controls its nuclear spin bath},\ }\href
  {https://doi.org/10.1103/PhysRevLett.105.216803} {\bibfield  {journal}
  {\bibinfo  {journal} {Phys.\ Rev.\ Lett.}\ }\textbf {\bibinfo {volume}
  {105}},\ \bibinfo {pages} {216803} (\bibinfo {year}
  {2010}{\natexlab{a}})}\BibitemShut {NoStop}%
\bibitem [{\citenamefont {Reilly}\ \emph {et~al.}(2008)\citenamefont {Reilly},
  \citenamefont {Taylor}, \citenamefont {Laird}, \citenamefont {Petta},
  \citenamefont {Marcus}, \citenamefont {Hanson},\ and\ \citenamefont
  {Gossard}}]{Reilly_PRL08}%
  \BibitemOpen
  \bibfield  {author} {\bibinfo {author} {\bibfnamefont {D.~J.}\ \bibnamefont
  {Reilly}}, \bibinfo {author} {\bibfnamefont {J.~M.}\ \bibnamefont {Taylor}},
  \bibinfo {author} {\bibfnamefont {E.~A.}\ \bibnamefont {Laird}}, \bibinfo
  {author} {\bibfnamefont {J.~R.}\ \bibnamefont {Petta}}, \bibinfo {author}
  {\bibfnamefont {C.~M.}\ \bibnamefont {Marcus}}, \bibinfo {author}
  {\bibfnamefont {M.~P.}\ \bibnamefont {Hanson}},\ and\ \bibinfo {author}
  {\bibfnamefont {A.~C.}\ \bibnamefont {Gossard}},\ }\bibfield  {title}
  {\bibinfo {title} {Measurement of temporal correlations of the overhauser
  field in a double quantum dot},\ }\href
  {https://doi.org/10.1103/PhysRevLett.101.236803} {\bibfield  {journal}
  {\bibinfo  {journal} {Phys.\ Rev.\ Lett.}\ }\textbf {\bibinfo {volume}
  {101}},\ \bibinfo {pages} {236803} (\bibinfo {year} {2008})}\BibitemShut
  {NoStop}%
\bibitem [{\citenamefont {Malinowski}\ \emph
  {et~al.}(2017{\natexlab{a}})\citenamefont {Malinowski}, \citenamefont
  {Martins}, \citenamefont {Cywi{\'n}ski}, \citenamefont {Rudner},
  \citenamefont {Nissen}, \citenamefont {Fallahi}, \citenamefont {Gardner},
  \citenamefont {Manfra}, \citenamefont {Marcus},\ and\ \citenamefont
  {Kuemmeth}}]{Malinowski_PRL17}%
  \BibitemOpen
  \bibfield  {author} {\bibinfo {author} {\bibfnamefont {F.~K.}\ \bibnamefont
  {Malinowski}}, \bibinfo {author} {\bibfnamefont {F.}~\bibnamefont {Martins}},
  \bibinfo {author} {\bibfnamefont {{\L}.}~\bibnamefont {Cywi{\'n}ski}},
  \bibinfo {author} {\bibfnamefont {M.~S.}\ \bibnamefont {Rudner}}, \bibinfo
  {author} {\bibfnamefont {P.~D.}\ \bibnamefont {Nissen}}, \bibinfo {author}
  {\bibfnamefont {S.}~\bibnamefont {Fallahi}}, \bibinfo {author} {\bibfnamefont
  {G.~C.}\ \bibnamefont {Gardner}}, \bibinfo {author} {\bibfnamefont {M.~J.}\
  \bibnamefont {Manfra}}, \bibinfo {author} {\bibfnamefont {C.~M.}\
  \bibnamefont {Marcus}},\ and\ \bibinfo {author} {\bibfnamefont
  {F.}~\bibnamefont {Kuemmeth}},\ }\bibfield  {title} {\bibinfo {title}
  {Spectrum of the nuclear environment for gaas spin qubits},\ }\href
  {https://doi.org/10.1103/PhysRevLett.118.177702} {\bibfield  {journal}
  {\bibinfo  {journal} {Phys.\ Rev.\ Lett.}\ }\textbf {\bibinfo {volume}
  {118}},\ \bibinfo {pages} {177702} (\bibinfo {year}
  {2017}{\natexlab{a}})}\BibitemShut {NoStop}%
\bibitem [{\citenamefont {Witzel}\ and\ \citenamefont {{Das
  Sarma}}(2006)}]{Witzel_PRB06}%
  \BibitemOpen
  \bibfield  {author} {\bibinfo {author} {\bibfnamefont {W.~M.}\ \bibnamefont
  {Witzel}}\ and\ \bibinfo {author} {\bibfnamefont {S.}~\bibnamefont {{Das
  Sarma}}},\ }\bibfield  {title} {\bibinfo {title} {Quantum theory for electron
  spin decoherence induced by nuclear spin dynamics in semiconductor quantum
  computer architectures: Spectral diffusion of localized electron spins in the
  nuclear solid-state environment},\ }\href
  {https://doi.org/10.1103/PhysRevB.74.035322} {\bibfield  {journal} {\bibinfo
  {journal} {Phys.\ Rev.\ B}\ }\textbf {\bibinfo {volume} {74}},\ \bibinfo
  {pages} {035322} (\bibinfo {year} {2006})}\BibitemShut {NoStop}%
\bibitem [{\citenamefont {Witzel}\ and\ \citenamefont {{Das
  Sarma}}(2007)}]{Witzel_PRL07}%
  \BibitemOpen
  \bibfield  {author} {\bibinfo {author} {\bibfnamefont {W.~M.}\ \bibnamefont
  {Witzel}}\ and\ \bibinfo {author} {\bibfnamefont {S.}~\bibnamefont {{Das
  Sarma}}},\ }\bibfield  {title} {\bibinfo {title} {Multiple-pulse coherence
  enhancement of solid state spin qubits},\ }\href
  {https://doi.org/10.1103/PhysRevLett.98.077601} {\bibfield  {journal}
  {\bibinfo  {journal} {Phys.\ Rev.\ Lett.}\ }\textbf {\bibinfo {volume}
  {98}},\ \bibinfo {pages} {077601} (\bibinfo {year} {2007})}\BibitemShut
  {NoStop}%
\bibitem [{\citenamefont {Bluhm}\ \emph
  {et~al.}(2010{\natexlab{b}})\citenamefont {Bluhm}, \citenamefont {Foletti},
  \citenamefont {Neder}, \citenamefont {Rudner}, \citenamefont {Mahalu},
  \citenamefont {Umansky},\ and\ \citenamefont {Yacoby}}]{Bluhm_NatPhys11}%
  \BibitemOpen
  \bibfield  {author} {\bibinfo {author} {\bibfnamefont {H.}~\bibnamefont
  {Bluhm}}, \bibinfo {author} {\bibfnamefont {S.}~\bibnamefont {Foletti}},
  \bibinfo {author} {\bibfnamefont {I.}~\bibnamefont {Neder}}, \bibinfo
  {author} {\bibfnamefont {M.}~\bibnamefont {Rudner}}, \bibinfo {author}
  {\bibfnamefont {D.}~\bibnamefont {Mahalu}}, \bibinfo {author} {\bibfnamefont
  {V.}~\bibnamefont {Umansky}},\ and\ \bibinfo {author} {\bibfnamefont
  {A.}~\bibnamefont {Yacoby}},\ }\bibfield  {title} {\bibinfo {title} {Long
  coherence of electron spins coupled to a nuclear spin bath},\ }\href
  {https://doi.org/10.1038/nphys1856} {\bibfield  {journal} {\bibinfo
  {journal} {Nat. Phys.}\ }\textbf {\bibinfo {volume} {7}},\ \bibinfo {pages}
  {109} (\bibinfo {year} {2010}{\natexlab{b}})}\BibitemShut {NoStop}%
\bibitem [{\citenamefont {Medford}\ \emph {et~al.}(2012)\citenamefont
  {Medford}, \citenamefont {Cywi\'{n}ski}, \citenamefont {Barthel},
  \citenamefont {Marcus}, \citenamefont {Hanson},\ and\ \citenamefont
  {Gossard}}]{Medford_PRL12}%
  \BibitemOpen
  \bibfield  {author} {\bibinfo {author} {\bibfnamefont {J.}~\bibnamefont
  {Medford}}, \bibinfo {author} {\bibfnamefont {{\L}.}~\bibnamefont
  {Cywi\'{n}ski}}, \bibinfo {author} {\bibfnamefont {C.}~\bibnamefont
  {Barthel}}, \bibinfo {author} {\bibfnamefont {C.~M.}\ \bibnamefont {Marcus}},
  \bibinfo {author} {\bibfnamefont {M.~P.}\ \bibnamefont {Hanson}},\ and\
  \bibinfo {author} {\bibfnamefont {A.~C.}\ \bibnamefont {Gossard}},\
  }\bibfield  {title} {\bibinfo {title} {Scaling of dynamical decoupling for
  spin qubits},\ }\href {https://doi.org/10.1103/PhysRevLett.108.086802}
  {\bibfield  {journal} {\bibinfo  {journal} {Phys.\ Rev.\ Lett.}\ }\textbf
  {\bibinfo {volume} {108}},\ \bibinfo {pages} {086802} (\bibinfo {year}
  {2012})}\BibitemShut {NoStop}%
\bibitem [{\citenamefont {Malinowski}\ \emph
  {et~al.}(2017{\natexlab{b}})\citenamefont {Malinowski}, \citenamefont
  {Martins}, \citenamefont {Nissen}, \citenamefont {Barnes}, \citenamefont
  {Cywi\'n{}ski}, \citenamefont {Rudner}, \citenamefont {Fallahi},
  \citenamefont {Gardner}, \citenamefont {Manfra}, \citenamefont {Marcus},\
  and\ \citenamefont {Kuemmeth}}]{Malinowski_NatNano17}%
  \BibitemOpen
  \bibfield  {author} {\bibinfo {author} {\bibfnamefont {F.~K.}\ \bibnamefont
  {Malinowski}}, \bibinfo {author} {\bibfnamefont {F.}~\bibnamefont {Martins}},
  \bibinfo {author} {\bibfnamefont {P.~D.}\ \bibnamefont {Nissen}}, \bibinfo
  {author} {\bibfnamefont {E.}~\bibnamefont {Barnes}}, \bibinfo {author}
  {\bibfnamefont {{\L}.}~\bibnamefont {Cywi\'n{}ski}}, \bibinfo {author}
  {\bibfnamefont {M.~S.}\ \bibnamefont {Rudner}}, \bibinfo {author}
  {\bibfnamefont {S.}~\bibnamefont {Fallahi}}, \bibinfo {author} {\bibfnamefont
  {G.~C.}\ \bibnamefont {Gardner}}, \bibinfo {author} {\bibfnamefont {M.~J.}\
  \bibnamefont {Manfra}}, \bibinfo {author} {\bibfnamefont {C.~M.}\
  \bibnamefont {Marcus}},\ and\ \bibinfo {author} {\bibfnamefont
  {F.}~\bibnamefont {Kuemmeth}},\ }\bibfield  {title} {\bibinfo {title} {Notch
  filtering the nuclear environment of a spin qubit},\ }\href
  {https://doi.org/10.1038/nnano.2016.170} {\bibfield  {journal} {\bibinfo
  {journal} {Nature Nanotechnology}\ }\textbf {\bibinfo {volume} {12}},\
  \bibinfo {pages} {16} (\bibinfo {year} {2017}{\natexlab{b}})}\BibitemShut
  {NoStop}%
\bibitem [{\citenamefont {Makhlin}\ and\ \citenamefont
  {Shnirman}(2004)}]{Makhlin_PRL04}%
  \BibitemOpen
  \bibfield  {author} {\bibinfo {author} {\bibfnamefont {Y.}~\bibnamefont
  {Makhlin}}\ and\ \bibinfo {author} {\bibfnamefont {A.}~\bibnamefont
  {Shnirman}},\ }\bibfield  {title} {\bibinfo {title} {Dephasing of solid-state
  qubits at optimal points},\ }\href
  {https://doi.org/10.1103/PhysRevLett.92.178301} {\bibfield  {journal}
  {\bibinfo  {journal} {Phys. Rev. Lett.}\ }\textbf {\bibinfo {volume} {92}},\
  \bibinfo {pages} {178301} (\bibinfo {year} {2004})}\BibitemShut {NoStop}%
\bibitem [{\citenamefont {Bergli}\ \emph {et~al.}(2006)\citenamefont {Bergli},
  \citenamefont {Galperin},\ and\ \citenamefont {Altshuler}}]{Bergli_PRB06}%
  \BibitemOpen
  \bibfield  {author} {\bibinfo {author} {\bibfnamefont {J.}~\bibnamefont
  {Bergli}}, \bibinfo {author} {\bibfnamefont {Y.~M.}\ \bibnamefont
  {Galperin}},\ and\ \bibinfo {author} {\bibfnamefont {B.~L.}\ \bibnamefont
  {Altshuler}},\ }\bibfield  {title} {\bibinfo {title} {Decoherence of a qubit
  by a non-gaussian noise at an arbitrary working point},\ }\href
  {https://doi.org/10.1103/PhysRevB.74.024509} {\bibfield  {journal} {\bibinfo
  {journal} {Phys.\ Rev.\ B}\ }\textbf {\bibinfo {volume} {74}},\ \bibinfo
  {pages} {024509} (\bibinfo {year} {2006})}\BibitemShut {NoStop}%
\bibitem [{\citenamefont {Cywi{\'n}ski}(2014)}]{Cywinski_PRA14}%
  \BibitemOpen
  \bibfield  {author} {\bibinfo {author} {\bibfnamefont {{\L}.}~\bibnamefont
  {Cywi{\'n}ski}},\ }\bibfield  {title} {\bibinfo {title} {Dynamical-decoupling
  noise spectroscopy at an optimal working point of a qubit},\ }\href
  {https://doi.org/10.1103/PhysRevA.90.042307} {\bibfield  {journal} {\bibinfo
  {journal} {Phys. Rev. A}\ }\textbf {\bibinfo {volume} {90}},\ \bibinfo
  {pages} {042307} (\bibinfo {year} {2014})}\BibitemShut {NoStop}%
\bibitem [{cor()}]{correlations}%
  \BibitemOpen
  \bibinfo {note} {While the high frequency noise component in $\delta h$ is
  likely the result of charge noise, it is still reasonable to assume that the
  two noises are only weakly correlated, as detuning is mostly sensitive to
  electric field along the axis connecting the two dots, while the Overhauser
  fields are sensitive to all components of the noisy electric
  fields.}\BibitemShut {Stop}%
\bibitem [{\citenamefont {Cywi{\'n}ski}\ \emph {et~al.}(2008)\citenamefont
  {Cywi{\'n}ski}, \citenamefont {Lutchyn}, \citenamefont {Nave},\ and\
  \citenamefont {{Das Sarma}}}]{Cywinski_PRB08}%
  \BibitemOpen
  \bibfield  {author} {\bibinfo {author} {\bibfnamefont {{\L}.}~\bibnamefont
  {Cywi{\'n}ski}}, \bibinfo {author} {\bibfnamefont {R.~M.}\ \bibnamefont
  {Lutchyn}}, \bibinfo {author} {\bibfnamefont {C.~P.}\ \bibnamefont {Nave}},\
  and\ \bibinfo {author} {\bibfnamefont {S.}~\bibnamefont {{Das Sarma}}},\
  }\bibfield  {title} {\bibinfo {title} {How to enhance dephasing time in
  superconducting qubits},\ }\href {https://doi.org/10.1103/PhysRevB.77.174509}
  {\bibfield  {journal} {\bibinfo  {journal} {Phys.\ Rev.\ B}\ }\textbf
  {\bibinfo {volume} {77}},\ \bibinfo {pages} {174509} (\bibinfo {year}
  {2008})}\BibitemShut {NoStop}%
\bibitem [{\citenamefont {Kubo}(1962)}]{Kubo}%
  \BibitemOpen
  \bibfield  {author} {\bibinfo {author} {\bibfnamefont {R.}~\bibnamefont
  {Kubo}},\ }\bibfield  {title} {\bibinfo {title} {Generalized cumulant
  expansion method},\ }\href {https://doi.org/10.1143/JPSJ.17.1100} {\bibfield
  {journal} {\bibinfo  {journal} {J. Phys. Soc. Jpn.}\ }\textbf {\bibinfo
  {volume} {17}},\ \bibinfo {pages} {1100} (\bibinfo {year}
  {1962})}\BibitemShut {NoStop}%
\bibitem [{SM()}]{SM}%
  \BibitemOpen
  \bibinfo {note} {See Supplemental Material at
  http://link.aps.org/Supplemental/}\BibitemShut {NoStop}%
\bibitem [{\citenamefont {Martins}\ \emph {et~al.}(2016)\citenamefont
  {Martins}, \citenamefont {Malinowski}, \citenamefont {Nissen}, \citenamefont
  {Barnes}, \citenamefont {Fallahi}, \citenamefont {Gardner}, \citenamefont
  {Manfra}, \citenamefont {Marcus},\ and\ \citenamefont
  {Kuemmeth}}]{Martins_PRL16}%
  \BibitemOpen
  \bibfield  {author} {\bibinfo {author} {\bibfnamefont {F.}~\bibnamefont
  {Martins}}, \bibinfo {author} {\bibfnamefont {F.~K.}\ \bibnamefont
  {Malinowski}}, \bibinfo {author} {\bibfnamefont {P.~D.}\ \bibnamefont
  {Nissen}}, \bibinfo {author} {\bibfnamefont {E.}~\bibnamefont {Barnes}},
  \bibinfo {author} {\bibfnamefont {S.}~\bibnamefont {Fallahi}}, \bibinfo
  {author} {\bibfnamefont {G.~C.}\ \bibnamefont {Gardner}}, \bibinfo {author}
  {\bibfnamefont {M.~J.}\ \bibnamefont {Manfra}}, \bibinfo {author}
  {\bibfnamefont {C.~M.}\ \bibnamefont {Marcus}},\ and\ \bibinfo {author}
  {\bibfnamefont {F.}~\bibnamefont {Kuemmeth}},\ }\bibfield  {title} {\bibinfo
  {title} {Noise suppression using symmetric exchange gates in spin qubits},\
  }\href {https://doi.org/10.1103/PhysRevLett.116.116801} {\bibfield  {journal}
  {\bibinfo  {journal} {Phys. Rev. Lett.}\ }\textbf {\bibinfo {volume} {116}},\
  \bibinfo {pages} {116801} (\bibinfo {year} {2016})}\BibitemShut {NoStop}%
\bibitem [{\citenamefont {Koppens}\ \emph {et~al.}(2007)\citenamefont
  {Koppens}, \citenamefont {Klauser}, \citenamefont {Coish}, \citenamefont
  {Nowack}, \citenamefont {Kouwenhoven}, \citenamefont {Loss},\ and\
  \citenamefont {Vandersypen}}]{Koppens_PRL07}%
  \BibitemOpen
  \bibfield  {author} {\bibinfo {author} {\bibfnamefont {F.~H.~L.}\
  \bibnamefont {Koppens}}, \bibinfo {author} {\bibfnamefont {D.}~\bibnamefont
  {Klauser}}, \bibinfo {author} {\bibfnamefont {W.~A.}\ \bibnamefont {Coish}},
  \bibinfo {author} {\bibfnamefont {K.~C.}\ \bibnamefont {Nowack}}, \bibinfo
  {author} {\bibfnamefont {L.~P.}\ \bibnamefont {Kouwenhoven}}, \bibinfo
  {author} {\bibfnamefont {D.}~\bibnamefont {Loss}},\ and\ \bibinfo {author}
  {\bibfnamefont {L.~M.~K.}\ \bibnamefont {Vandersypen}},\ }\bibfield  {title}
  {\bibinfo {title} {Universal phase shift and nonexponential decay of driven
  single-spin oscillations},\ }\href
  {https://doi.org/10.1103/PhysRevLett.99.106803} {\bibfield  {journal}
  {\bibinfo  {journal} {Phys.\ Rev.\ Lett.}\ }\textbf {\bibinfo {volume}
  {99}},\ \bibinfo {pages} {106803} (\bibinfo {year} {2007})}\BibitemShut
  {NoStop}%
\bibitem [{\citenamefont {Cywi{\'n}ski}\ \emph {et~al.}(2009)\citenamefont
  {Cywi{\'n}ski}, \citenamefont {Witzel},\ and\ \citenamefont {{Das
  Sarma}}}]{Cywinski_PRB09}%
  \BibitemOpen
  \bibfield  {author} {\bibinfo {author} {\bibfnamefont {{\L}.}~\bibnamefont
  {Cywi{\'n}ski}}, \bibinfo {author} {\bibfnamefont {W.~M.}\ \bibnamefont
  {Witzel}},\ and\ \bibinfo {author} {\bibfnamefont {S.}~\bibnamefont {{Das
  Sarma}}},\ }\bibfield  {title} {\bibinfo {title} {Pure quantum dephasing of a
  solid-state electron spin qubit in a large nuclear spin bath coupled by
  long-range hyperfine-mediated interaction},\ }\href
  {https://doi.org/10.1103/PhysRevB.79.245314} {\bibfield  {journal} {\bibinfo
  {journal} {Phys.\ Rev.\ B}\ }\textbf {\bibinfo {volume} {79}},\ \bibinfo
  {pages} {245314} (\bibinfo {year} {2009})}\BibitemShut {NoStop}%
\bibitem [{sem()}]{semilinked}%
  \BibitemOpen
  \bibinfo {note} {Note that semi-linked terms contribution, Eq.~(\ref{R2ktlf})
  can be substantial, providing additional weight to the perpendicular
  noise.}\BibitemShut {Stop}%
\end{thebibliography}%

\pagebreak
\onecolumngrid
\hspace*{-2cm}
\includegraphics{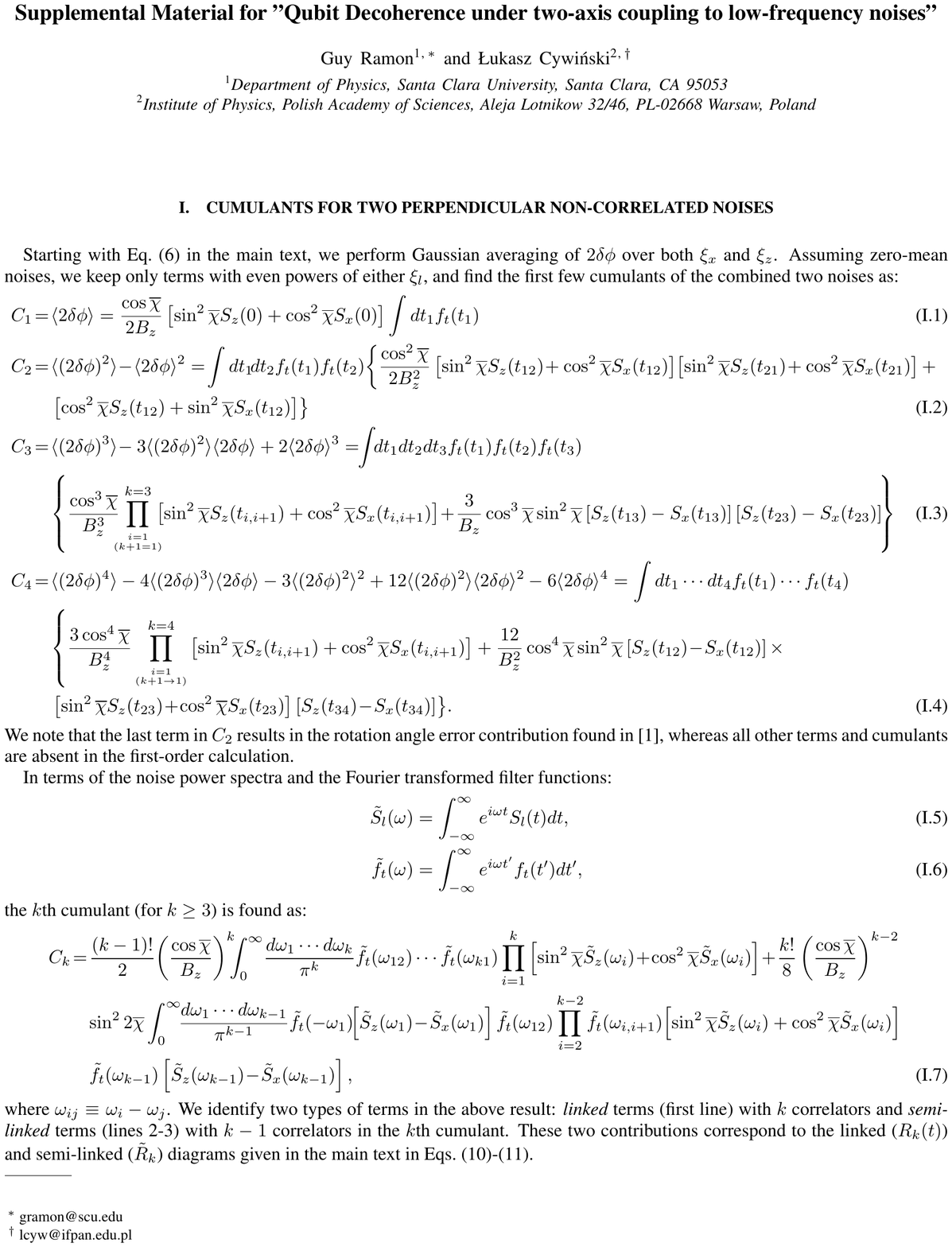}
\hspace*{-2cm}
\includegraphics{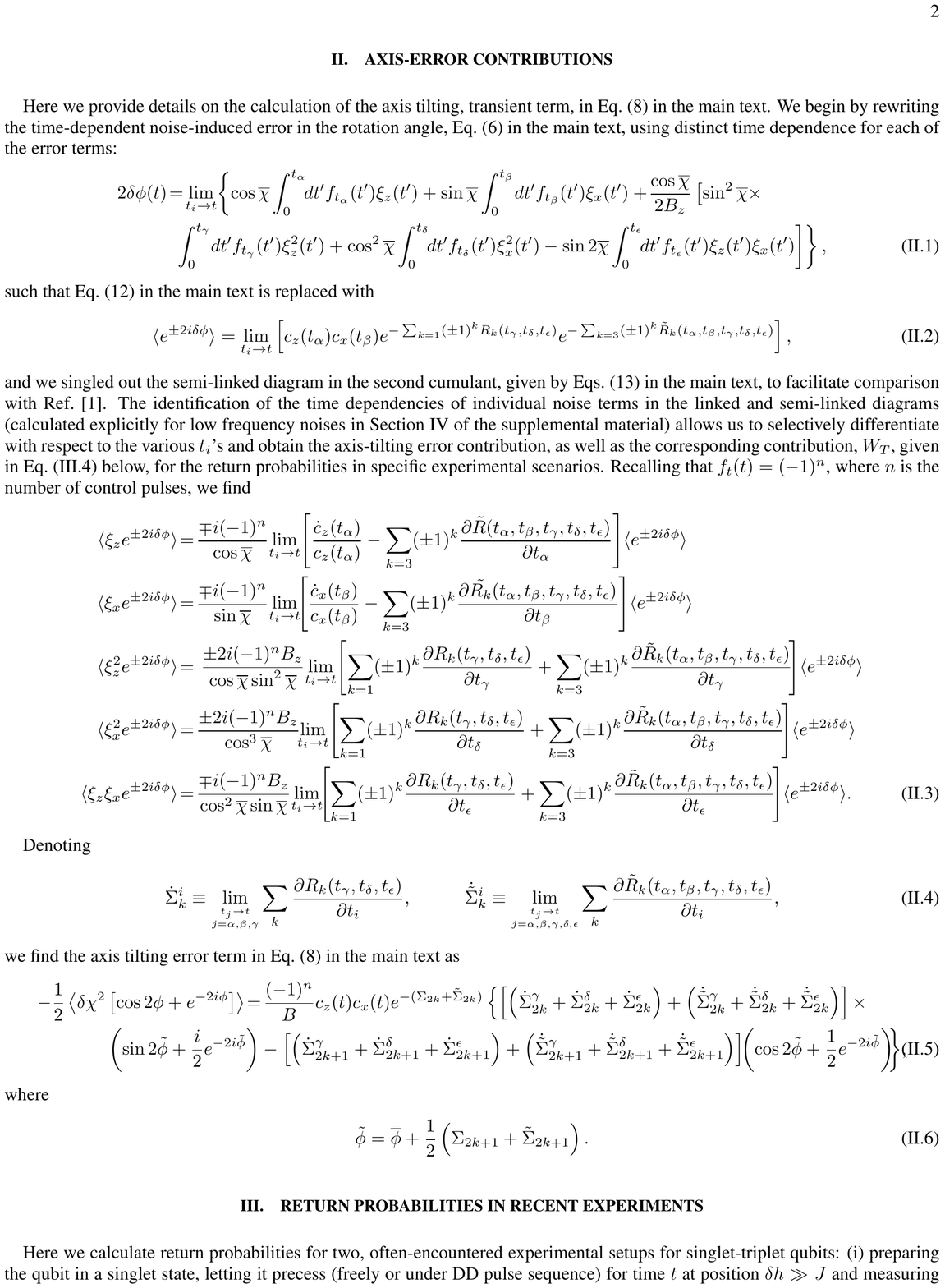}
\hspace*{-2cm}
\includegraphics{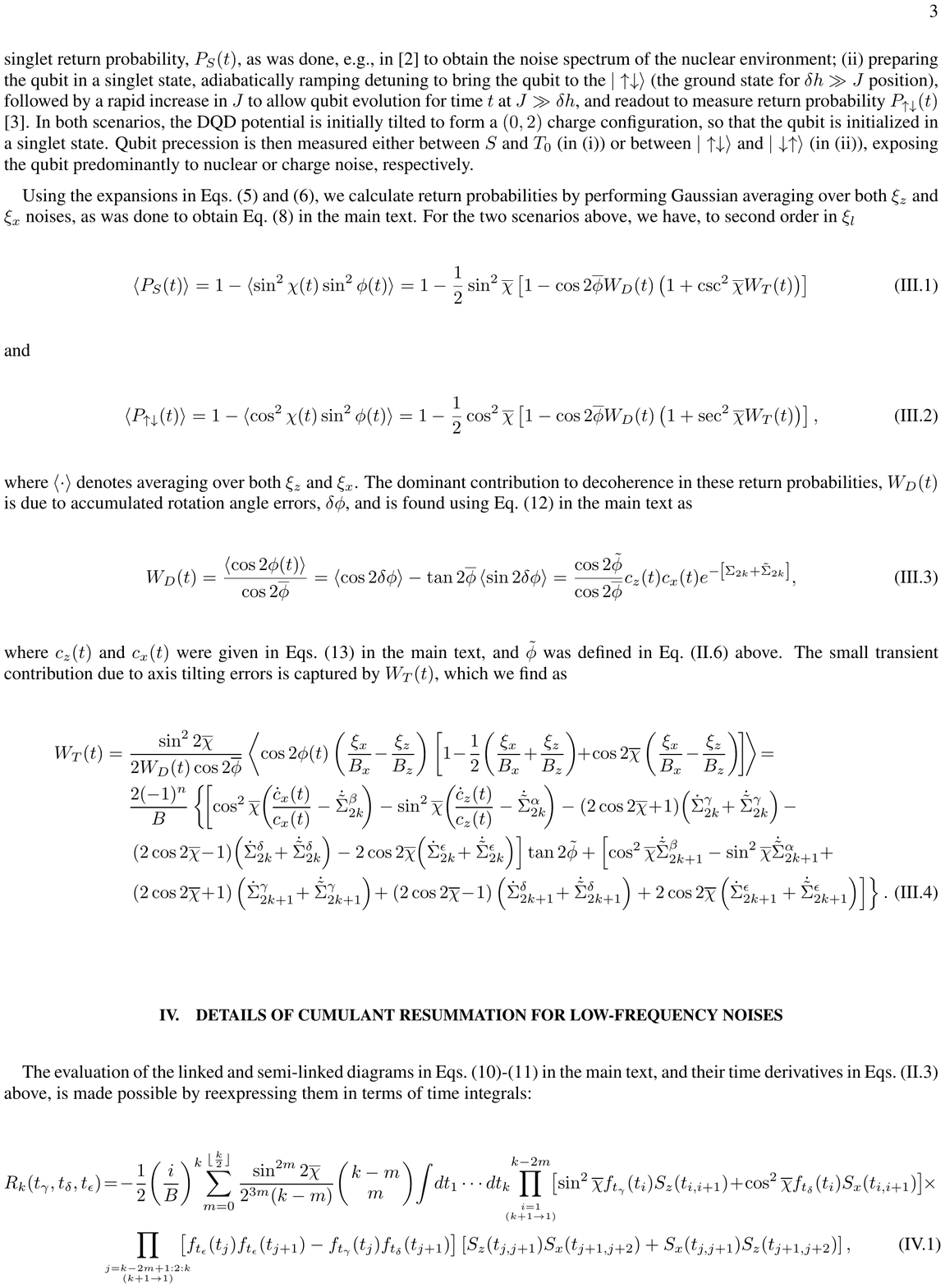}
\hspace*{-2cm}
\includegraphics{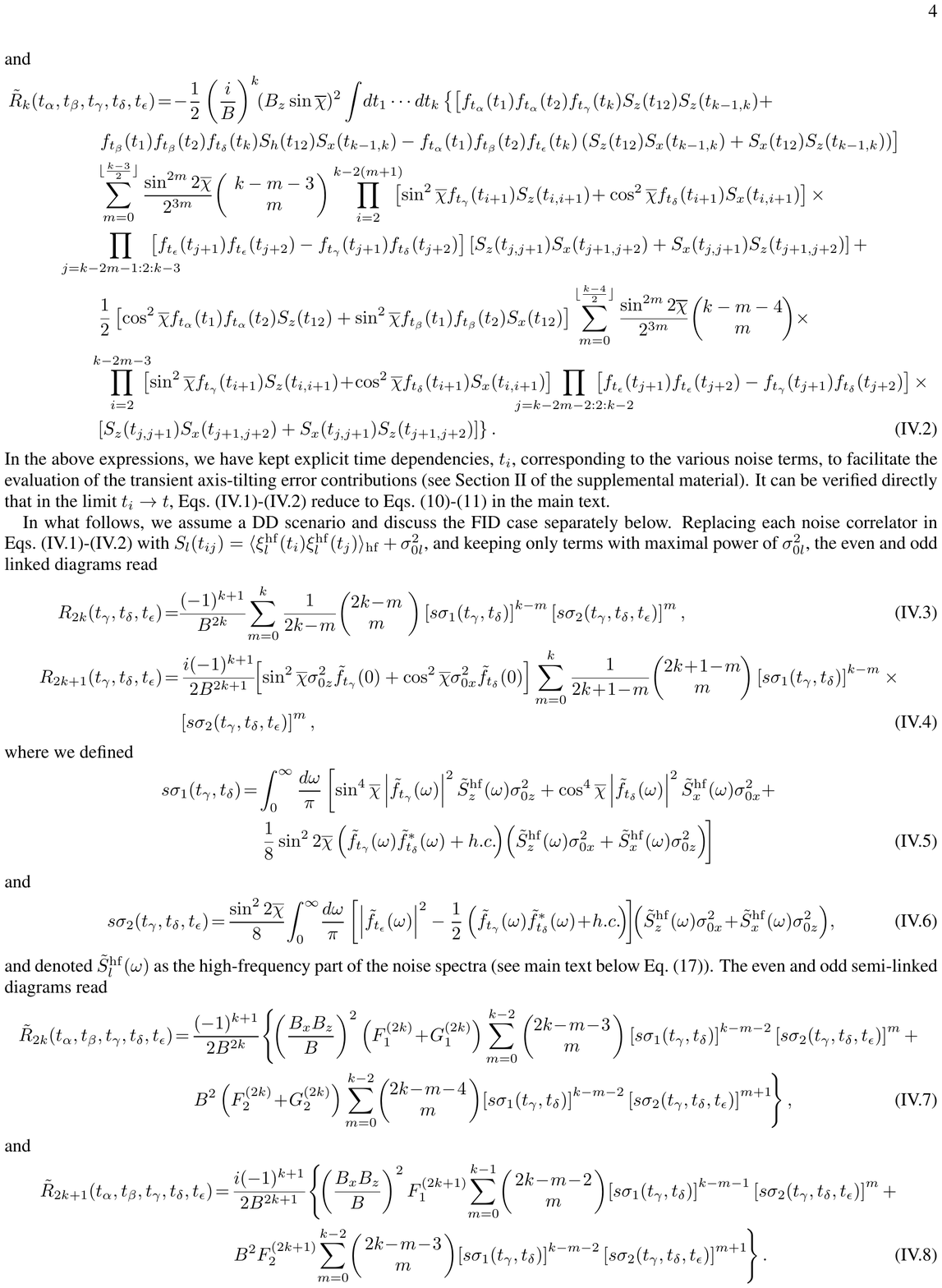}
\hspace*{-2cm}
\includegraphics{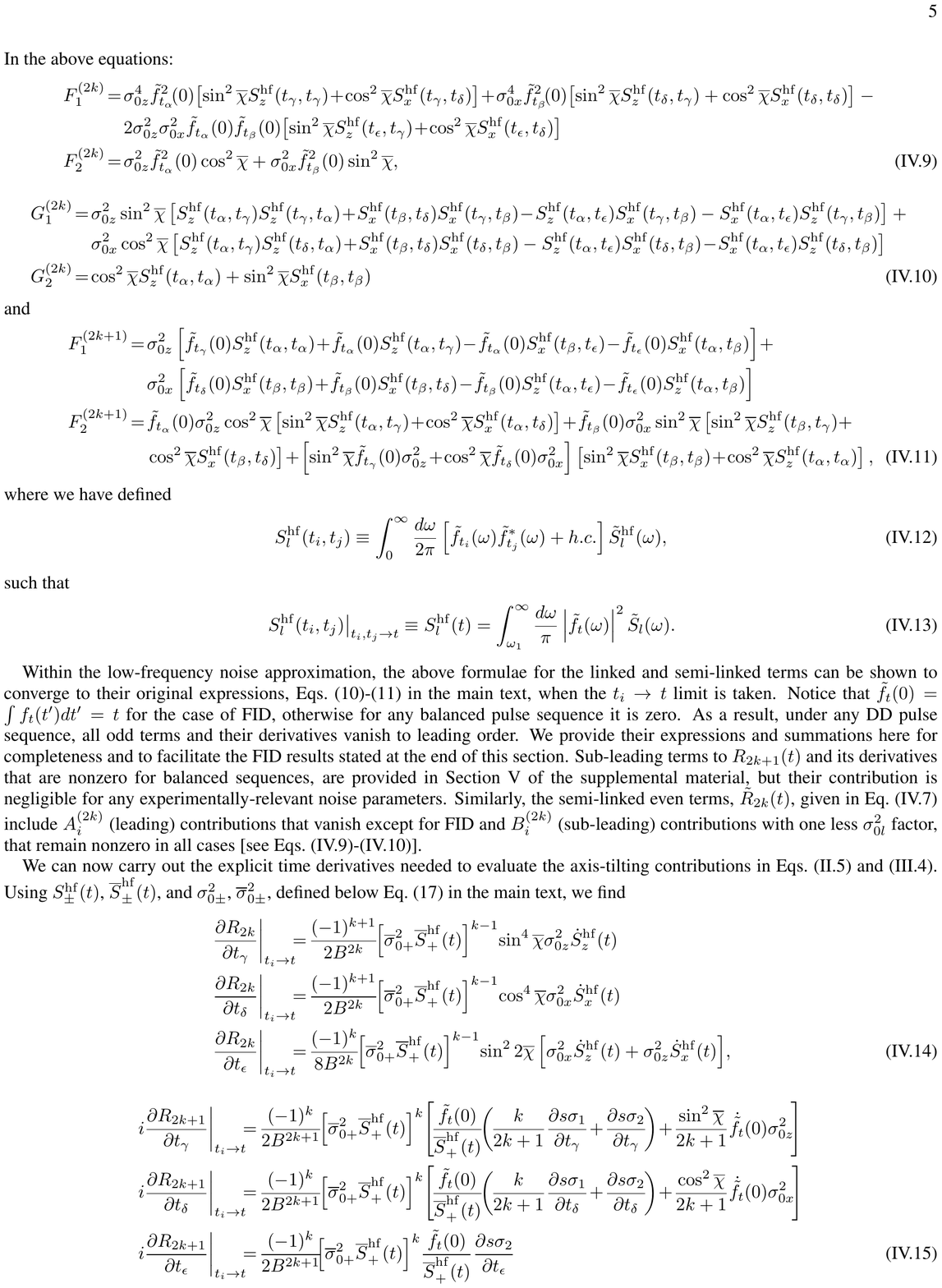}
\hspace*{-2cm}
\includegraphics{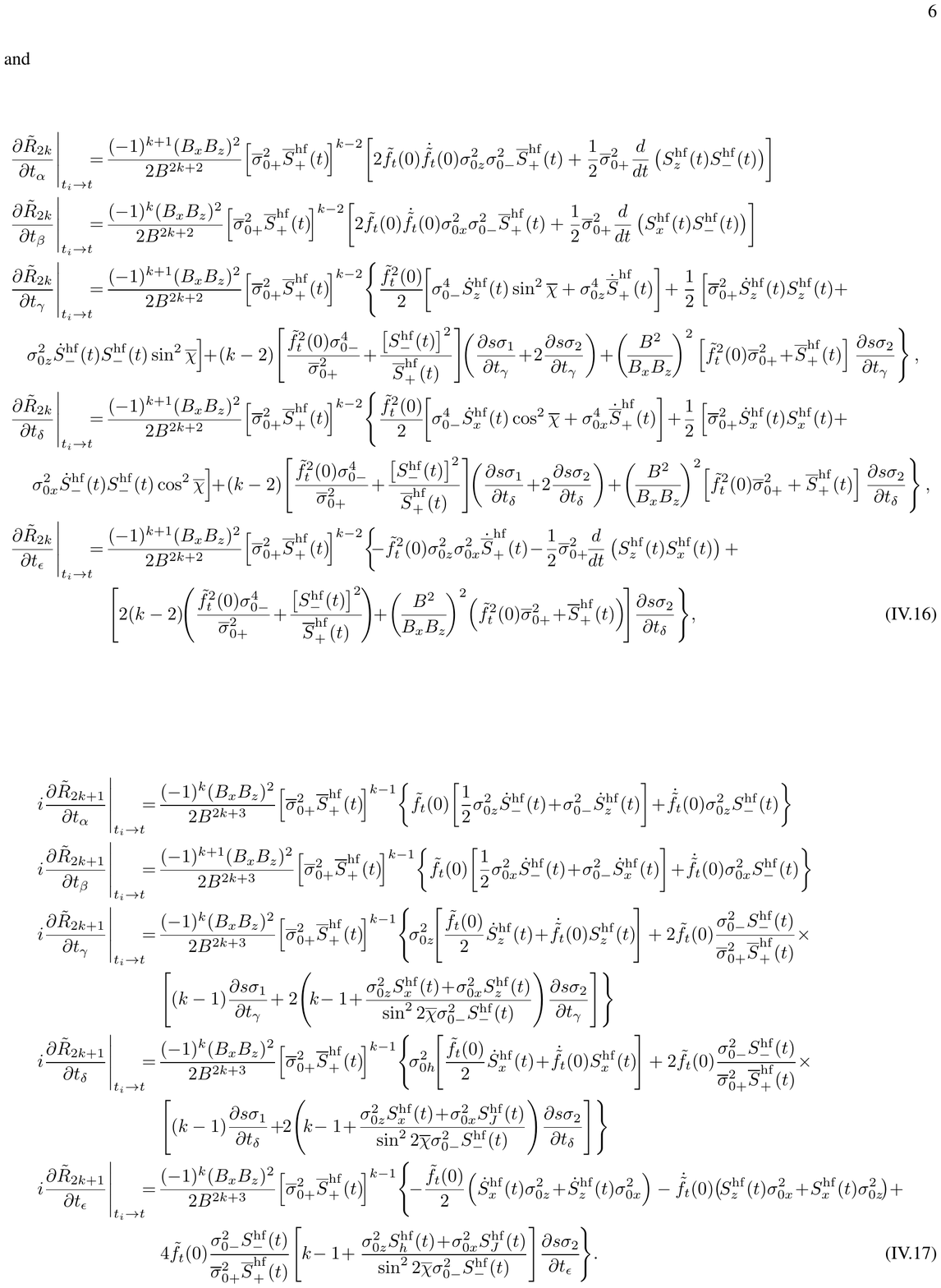}
\hspace*{-2cm}
\includegraphics{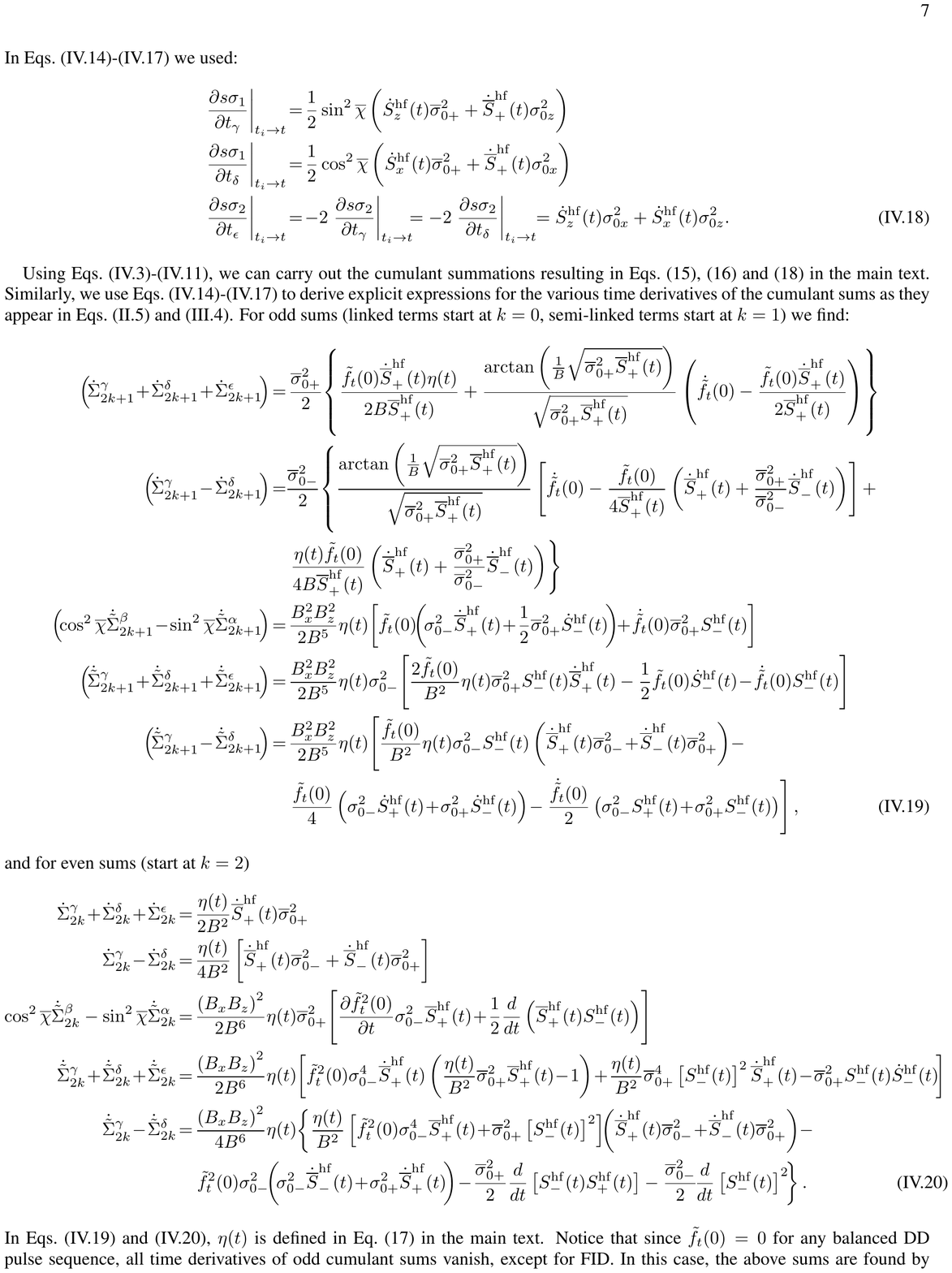}
\hspace*{-2cm}
\includegraphics{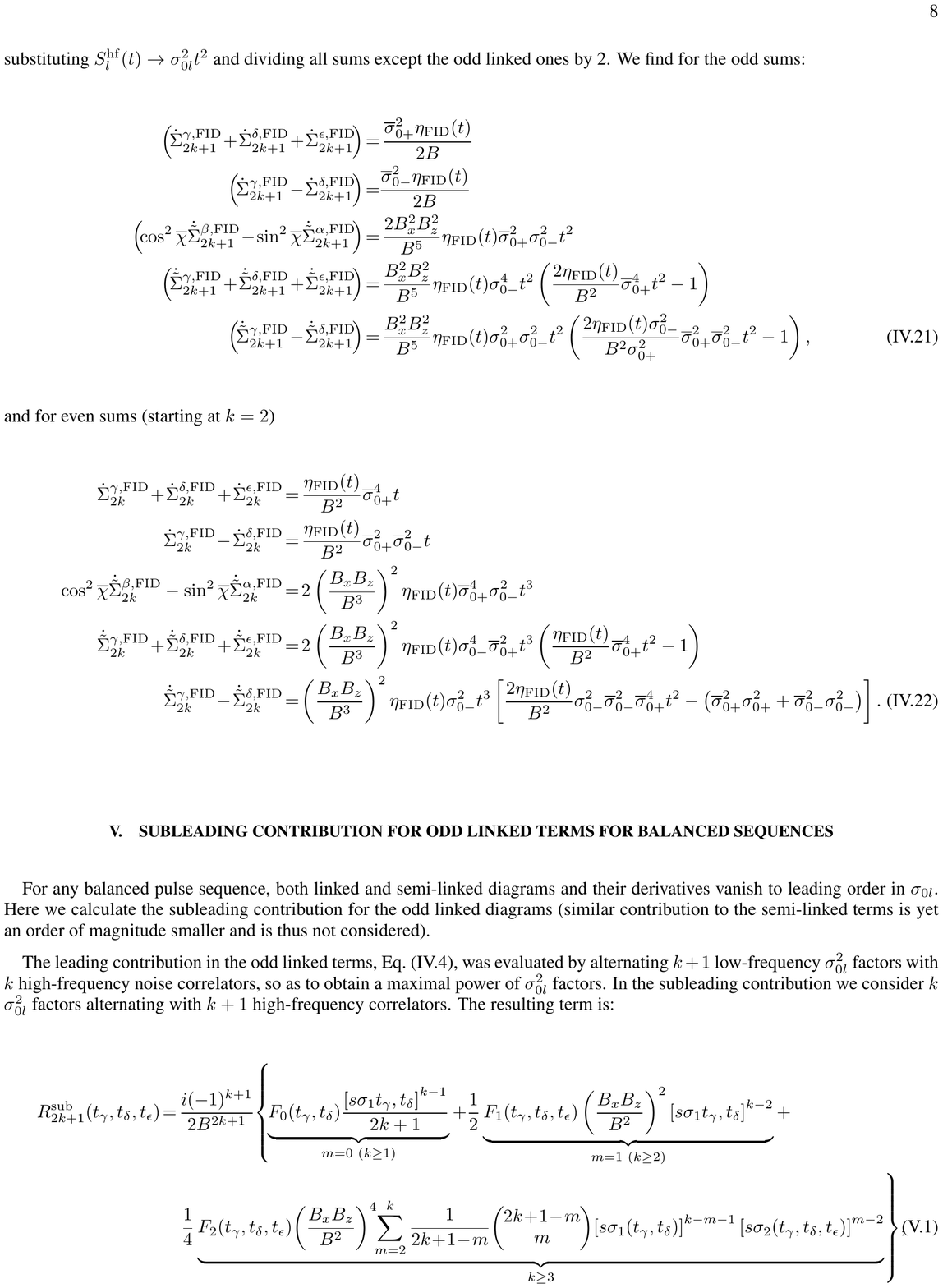}
\hspace*{-2cm}
\includegraphics{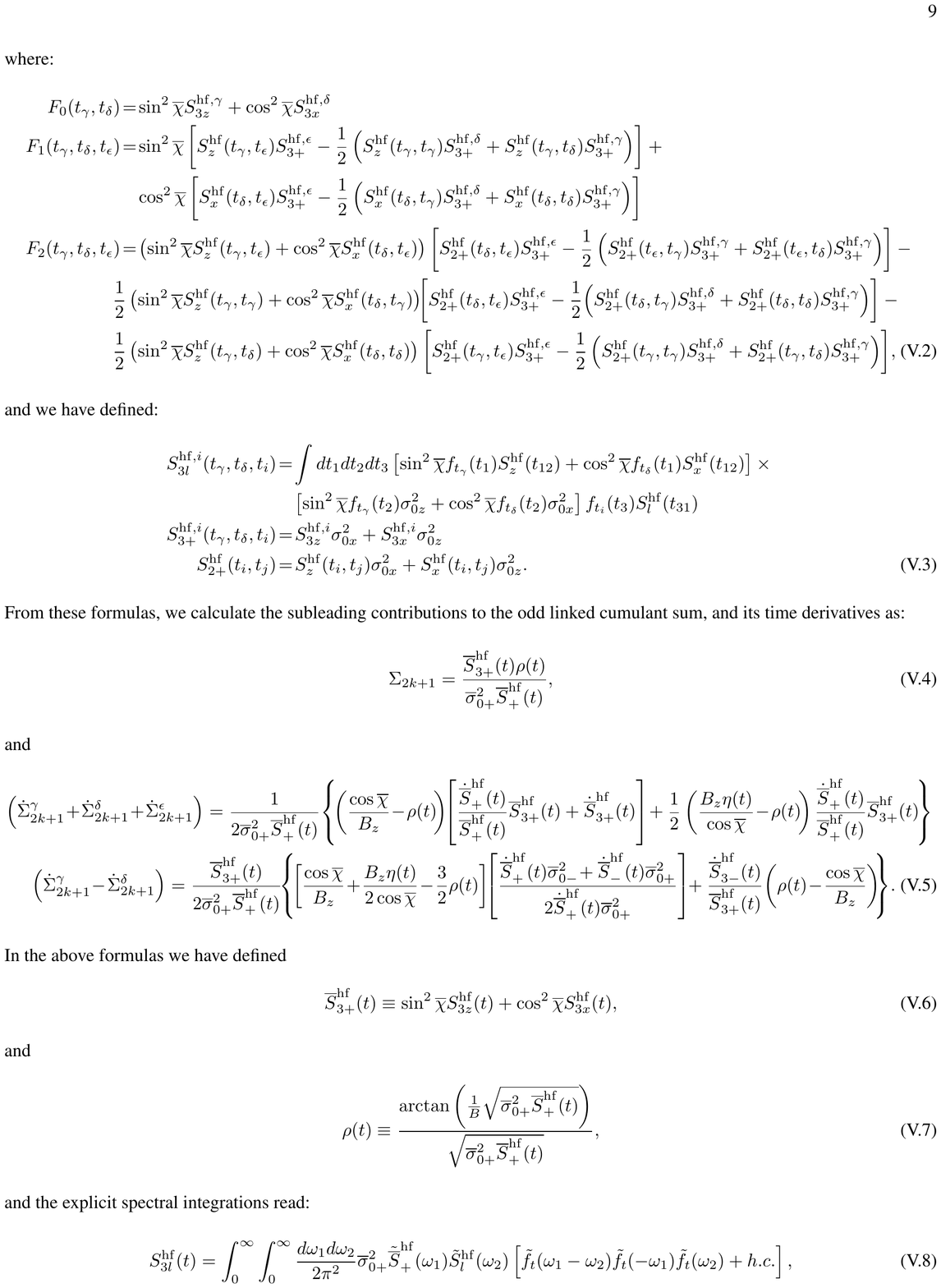}
\hspace*{-2cm}
\includegraphics{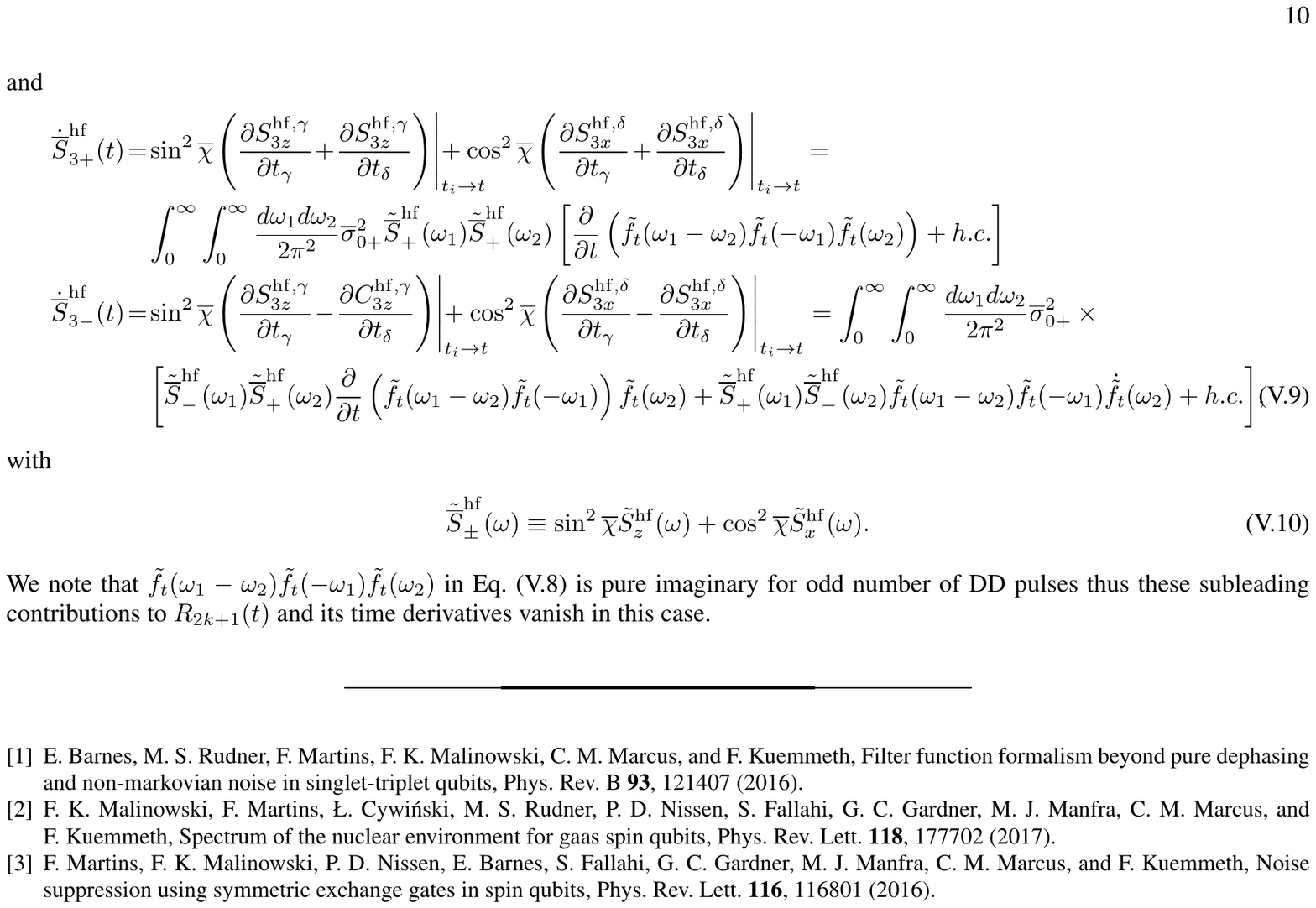}

\end{document}